\documentclass[journal,10pt,draftclsnofoot,onecolumn]{IEEEtran}
\usepackage[dvips]{graphicx}
\usepackage{amsmath,amssymb,color,verbatim}

\def\norm#1{\left\| #1 \right\|}
\newtheorem{definition}{Definition}[section]
\newtheorem{thm}{Theorem}[section]
\newtheorem{proposition}[thm]{Proposition}
\newtheorem{lemma}[thm]{Lemma}
\newtheorem{corollary}[thm]{Corollary}
\newtheorem{exam}{Example}[section]
\def\diag{\textrm{diag}}

\renewcommand{\_}{\underline}

\def\bea{\begin{IEEEeqnarray}{rCl}} 
\def\eea{\end{IEEEeqnarray}}
\def\beq{\begin{equation}}
\def\eeq{\end{equation}}
\def\bean{\begin{IEEEeqnarray*}{rCl}} 
\def\eean{\end{IEEEeqnarray*}} 

\newcommand{\const}[1]{\textnormal{\usefont{U}{eur}{m}{n}%
\selectfont #1}}

\newtheorem{remark}{Remark}[section]

\DeclareMathOperator*{\tr}{Tr}

\DeclareMathOperator*{\End}{End}
\DeclareMathOperator*{\Ker}{Ker}
\providecommand{\abs}[1]{\ensuremath{\left\lvert #1 \right\rvert}}
\providecommand{\norm}[1]{\ensuremath{\left\Vert #1 \right\Vert}}

\providecommand{\vv}[1]{\textquotedblleft #1\textquotedblright}
\newcommand{\Q}{\mathbb{Q}}
\newcommand{\Z}{\mathbb{Z}}
\newcommand{\C}{\mathbb{C}}
\newcommand{\R}{\mathbb{R}}
\newcommand{\Hh}{\mathbb{H}}

\DeclareMathOperator*{\ad}{ad}

\DeclareMathOperator*{\SL}{SL}

\DeclareMathOperator*{\SNR}{SNR}
\DeclareMathOperator*{\Vol}{Vol}
\providecommand{\abs}[1]{\ensuremath{\left\lvert #1 \right\rvert}}
\providecommand{\norm}[1]{\ensuremath{\left\Vert #1 \right\Vert}}

\providecommand{\vv}[1]{\textquotedblleft #1\textquotedblright}

\newcommand{\D}{{\mathcal D}}
\newcommand{\F}{{\mathbf F}}

\newcommand{\OO}{{\mathcal O}}

\newcommand{\A}{{\mathcal A}}

\def\SNR{\textrm{SNR}}

\begin{document}

\title{Inverse Determinant Sums  and Connections Between  Fading Channel Information Theory and Algebra 
\thanks{The research of R. Vehkalahti is funded by   Academy of Finland  grants  \#131745 and \#252457.}
\thanks{The research of L. Luzzi was funded in part by a Marie Curie Fellowship (FP7/2007-2013, grant agreement PIEF-GA-2010-274765).
}
\thanks{Part of this work appeared at ISIT 2011 \cite{VLISIT}, at ITW 2011 \cite{VLITW} and ISIT 2012 \cite{LV2012}.}
}

\author{Roope Vehkalahti,   Hsiao-feng (Francis) Lu, \emph{Member, IEEE}, Laura Luzzi, \emph{Member, IEEE}
\thanks{R. Vehkalahti is with the Department of Mathematics, FI-20014, University of Turku, Finland  (e-mail: roiive@utu.fi). During  part of this  work he was visiting  the Department of Mathematics, Chaire des structures alg\'ebriques et ge\'om\'etriques at   \'Ecole Polytechnique F\'ed\'erale de Lausanne.}
\thanks{H.-f. Lu is with the Department of Electrical Engineering, National Chiao Tung University, Hsinchu, Taiwan (e-mail:francis@mail.nctu.edu.tw). }
\thanks{L. Luzzi is with Laboratoire ETIS (ENSEA - Universit\'e de Cergy-Pontoise - CNRS), 95014 Cergy-Pontoise, France (e-mail:
laura.luzzi@ensea.fr). During part of this work she was with the Department of Electrical and Electronic Engineering, Imperial College London, London SW7 2AZ, United Kingdom.}

}


\maketitle

\begin{abstract}
This work concentrates on the study of inverse determinant sums, which
arise from the union bound on the error probability, as a tool for
designing and analyzing algebraic space-time block codes.
A general framework to study these sums is established, and the connection
between asymptotic growth of inverse determinant sums and the
diversity-multiplexing gain trade-off is investigated.
It is proven that the growth of the inverse determinant sum of a division
algebra-based space-time code is completely determined by the growth of
the unit group. This reduces the inverse determinant sum analysis to
studying certain asymptotic integrals in Lie groups. Using recent methods
from ergodic theory, a complete classification of the inverse determinant
sums of the most well known algebraic space-time codes is provided.
The approach reveals an interesting and tight relation between
diversity-multiplexing gain trade-off and point counting in Lie groups.

\end{abstract}

\begin{keywords} division algebra, space-time block codes (STBCs), multiple-input multiple-output (MIMO), unit group, zeta functions, diversity-multiplexing gain trade-off (DMT), algebra, number theory, Lie groups.\end{keywords}

\section{Introduction}

In this paper we introduce a new technique to analyze the performance of lattice space-time block codes in  the high SNR regime. 
By developing the analysis based on the union bound of the pairwise error probabilities of such codes, we show that the high-SNR performance is related to the asymptotic behavior of the inverse determinant sums of these codes.  
The new performance criterion based on inverse determinant sums fills in the middle ground between the Diversity-Multiplexing Trade-off (DMT) \cite{ZT} and the normalized minimum determinant.

The normalized minimum determinant criterion has been used effectively to choose which space-time code one should use in order to get the best performance. For a relatively high SNR level the optimization work has produced very good results. However, this criterion concentrates on minimizing the worst case pairwise error probability, and does not consider its overall distribution, disregarding for example the question of how many times the worst case scenario occurs. 

The DMT, on the other hand, is a measure that considers the overall error probability, but only in the asymptotic sense as the SNR and codebook size grow to infinity. Moreover, the DMT focuses only on the diversity exponent, and in many cases it is too coarse for practical code design. For example, from the DMT point of view almost all  full-rate division algebra-based codes are equivalent in diversity exponent, while their actual performances often differ strongly.
 
The asymptotic growth of the inverse determinant sum captures something between these two concepts. Our analysis reveals that the diversity-multiplexing gain bounds of Zheng and Tse \cite{ZT} constitute general lower bounds for the asymptotic growth of inverse determinant sums.  The bounds depend on the dimension of the lattice and on the number of transmit and receive antennas. Achieving such bounds immediately proves that a code is DMT optimal for multiplexing gains between [0,1], while in other cases the asymptotic growth provides information on the DMT for multiplexing gains in this region. 
Furthermore, the behavior of inverse determinant sums  can be analyzed with great accuracy and can provide information both on the normalized minimum determinant 
and DMT. 
But, in this paper we are mostly interested in the interplay between DMT and inverse determinant sums and will only consider exponents of the growth of the latter.

While the first part of the paper is about stating the problem and proving general lower bounds, the second part concentrates on analyzing the growth  of inverse determinant sums of large classes of algebraic space-time codes. Most of the division algebra-based codes are subsets of an \emph{order} \cite{HLL} inside the division algebra. Using orders guarantees the non-vanishing determinant property (NVD), which has been shown to be a sufficient  criterion for DMT optimality for lattice codes in the space $M_n(\C)$ having full rank $2n^2$ \cite{EKPKL} and \cite{TV}.

We will prove that the growth of the inverse determinant sum of a division algebra-based space-time code depends only on the asymptotic growth of the norms of the \emph{unit group} of the underlying order, and can be computed from invariants of the corresponding algebra. This allows us to give a complete analysis of the 
inverse determinant sums of the most commonly used division algebra-based space-time codes.

Maybe unsurprisingly, we find that for all the $2n^2$-dimensional division algebra-based codes, this growth corresponds exactly to the DMT lower bound. This offers an intuitive explanation of why these codes are DMT optimal and of why the simple normalized minimum determinant optimization has been so successful. However, when we consider division algebra-based lattice codes having less than full rank in $M_n(\C)$, we will see that the choice of the algebra can have a dramatic effect on the growth of the inverse determinant sum. As we will see in Subsection \ref{example}, different growth rates seem to lead also to vast differences in performance. Our results thus provide a general framework to compare the DMTs of different types of algebraic space-time code constructions.

While our analysis of division algebra codes relies on algebraic 
concepts such as the \emph{Dedekind} and \emph{Hey} zeta functions as well as on the analysis of unit group, 
 our work is fundamentally based on recent results in the field of \emph{ergodic theory}.
The reason that we are able to analyze the asymptotic behavior of the norms of the unit group, is that this group can be seen as a \emph{lattice} inside a \emph{Lie group}, and the asymptotic growth problem is related to a \emph{point counting problem for Lie groups}. 

The study of such point counting problems
is part of a rather recent but highly developed mathematical area having a rich spectrum of general methods.  For the most recent approach based on ergodic methods we refer to the monograph by Gorodnik and Nevo \cite{GN}.

We point out the surprising tightness of the relation between algebraic and information-theoretic results. In some cases the completely general lower bounds for inverse determinant sums, derived from information theory, do meet the upper bounds derived
from deep algebraic results. In the case of complex quadratic center, the DMT results manage to correctly predict the distribution of (algebraic) norms of elements of an order in a division algebra.

\subsection{Contents of the paper}
We begin by recalling the notion of DMT and some basic definitions of lattice theory. In Section \ref{sec:3} we  first formalize the inverse determinant sum problem, give an example of 
its practical interest as well as
some simple bounds for the asymptotic growth.  We then consider how the asymptotic behavior of the inverse determinant sum of 
a space-time code is related to its DMT. As an example, we study the determinant sum for the Alamouti code \cite{Alam} and recognize that it is the truncated Epstein zeta function. This gives a new proof of the fact that the Alamouti code is DMT optimal for a single-antenna receiver.  Finally in Section \ref{mathematicalcomment} we point out how the DMT results can help to study some problems arising from lattice theory. 

In Section \ref{sec:4} we study diagonal MISO codes from algebraic number fields. We show how the corresponding inverse determinant sum can be asymptotically approximated by combining the information about the geometric structure of the unit group and about the behavior of the truncated {\em Dedekind zeta function} at integer points. This study reveals that  the growth of the inverse determinant sums of different number field codes, coming from fields with equal degree,  only differ by a constant term.  As a corollary we give a new proof of the DMT-optimality of these algebraic codes. In order to keep the presentation of the paper suitable for a larger audience we have postponed some of the proofs to Section \ref{sec:proof}.

In Section \ref{sec:6} we begin to study inverse determinant sums of division algebra-based space-time codes. First, we show how these inverse determinant sums depend on the behavior of the \emph{Hey zeta function} and of the unit group of an order of the algebra. In particular we prove that the growth of the inverse determinant sum depends only on the algebraic properties of the division algebra and in particular on the unit group.

In Section \ref{sec:7} we translate the  inverse determinant sums results to the the language of DMT and give new DMT lower bounds for a large class of division algebra-based codes.

Section  \ref{lie_sec} is devoted to the point counting problem in Lie groups.  Results of asymptotic growth rate are given for discrete lattice subgroups of three Lie groups that are most central to our theory. After arming ourselves with enough point counting results,  we will give the proofs of  Section \ref{sec:6} in Section \ref{divsumsecproofs}.

Finally we have collected some relevant Lie algebra theory, that is needed in Section \ref{sec:7} in the Appendix.

We have tried to keep most of the paper easily approachable. Apart from Section \ref{sec:proof}, the first seven sections should be readable with a rather modest algebraic background.

\subsection{Related work}
The study of inverse determinant sums is a natural question in multiple antenna fading channels. For example, in \cite{TV}, Tavildar and Viswanath analyzed the DMT of several coding schemes by using the union bound approach. However, they  did not consider determinant sums, but eventually restricted their attention to coding schemes where elementary combinatorial methods could be applied. In  \cite{RVCC} the authors studied the blind detection of QAM and PAM symbols. In their analysis they considered the Dedekind zeta function of the field $\Q(i)$. In Example \ref{QAM-zeta} we discuss briefly how their approach can be seen as the most simple case of our theory.   

Already in 1998 Boutros and Viterbo considered the \emph{product kissing number} in the context of number field codes \cite{BV}, and noted that one should develop a criterion which could take into account not only the minimum determinant, but also the multiplicity of occurrence of the worst case scenario. The normalized criterion presented in the beginning of Section \ref{basicprob} addresses this issue (and more). 
As presented in Section \ref{constremark}, our rough asymptotic methods can be straightforwardly modified to work in the way Boutros  and Viterbo probably had in mind. For a recent work on product kissing numbers we  refer the reader to \cite{WZ}, where the authors consider  this question  in the context of \emph{quasi-orthogonal codes}.

The closest and independent line of research that is  related to  our work 
has been carried out recently by  F. Oggier and J.-C. Belfiore.  In \cite{BOICC} they consider Rayleigh fast fading  wiretap channels and number field codes. In particular by measuring  error probabilities in wiretap channels they end up with the same number field sums as we do. In \cite{BO} Belfiore and Oggier consider the Rayleigh fading MIMO wiretap channel, where their work also leads to the same inverse determinant sums.  However, their analysis considers only the Alamouti code.

In the crossroad of ours and the work of Oggier and Belfiore is the work by Hollanti and Viterbo \cite{HV}. They considered the error probability of wiretap codes using similar methods to ours. In particular their goal has been to give a finite version of the bound  given in Section \ref{constremark}.

 While the growth of inverse determinant sums of orders of division  algebras or rings of algebraic integers are related to distribution of norms of elements in these rings, to  the best of our knowledge, there doesn't seem to be any previous algebraic work on the subject.

\subsection{Main contributions of this paper}

The contributions of this paper are the following.
\begin{itemize}

\item A formal definition of inverse determinant sums as a code design criterion and a tool for analyzing DMT of a code.
\item General upper and lower bounds for inverse determinant sums.
\item A connection between error probability, Dedekind zeta function and unit group of algebraic number field codes.
\item A connection among error probability, Hey zeta function and unit group of division algebra codes.
\item A complete analysis of the growth of inverse determinant sums of several families of algebraic space-time codes.
\item New DMT lower bounds for the aforementioned division algebra codes.
\end{itemize}

\section{The Players}

\subsection{The DMT}

Consider a Rayleigh block fading MIMO channel with $n_t$ transmit and $n_r$ receive antennas. The channel is assumed to be fixed for a block of $T$ channel uses, but vary in an independent and identically distributed (i.i.d.) fashion from one block to another. Thus, the channel input-output relation can be written as 
\beq
Y=\sqrt{\frac{\rho}{n_t}}HX +N, \label{eq:channel}
\eeq
where $H \in M_{n_r \times n_t}(\C)$ is the channel matrix and $N\in M_{n_r \times T}(\C) $ is the noise matrix. The entries of $H$ and $N$ are assumed to be  i.i.d. zero-mean complex circular symmetric Gaussian random variables with variance 1. $X \in M_{n_t\times T}(\C)$ is the transmitted codeword, and $\rho$ denotes the signal-to-noise ratio (SNR). 

Assuming the channel is block-ergodic, and matrix $H$ is known completely to the receiver but not to the transmitter, Telatar \cite{Tel} showed that the capacity of the MIMO channel \eqref{eq:channel} is given by
\bea
{\const C}(\rho) &=& {\mathbb E} \log \det \left( I_{n_r} + \frac{\rho}{n_t} H H^\dag \right)\nonumber\\
&=& \min\{ n_t, n_r\} \log \rho + O(1), \label{eq:capacity}
\eea
in bits per channel use (bpcu), provided that the transmitted codeword $X$ satisfies an average power constraint ${\mathbb E} \norm{X}_F^2 \leq T n_t$. The logarithm in \eqref{eq:capacity} is taken with base 2. 

The capacity formula \eqref{eq:capacity} means that an error-free communication, i.e., having an error probability arbitrarily close to 0, over the MIMO channel \eqref{eq:channel} is possible only when transmission rate $R \leq {\const C}(\rho)$. However, for any fixed SNR level $\rho$, it is commonly believed that making the error probability arbitrarily small requires a coded transmission over infinitely many blocks of channel, which is by no means practical. As a result, it is of a great interest to determine how small the error probability can be when the coding is limited to only one block of $T$ channel uses. This has been studied in great detail by Zheng and Tse in \cite{ZT}. Below we provide a brief overview of some of the important results in \cite{ZT}, including the notion of DMT. 

\begin{definition}\label{def:stbc}
A {\em space-time block code} (STBC) $C$  for some designated SNR level is a set of $n_t \times T$ complex matrices satisfying the following average power constraint
\begin{equation}\label{powernorm}
\frac{1}{\abs{C}}\sum_{X \in C} \norm{X}_F^2 \leq T n_t. 
\end{equation}
The rate of code $C$ is $R=\frac{1}{T} \log \abs{C}$ in bpcu. A coding scheme $\{ C(\rho)\}$ of STBC is a family of STBCs, one at each SNR level. The rate for code $C(\rho)$ is thus $R(\rho)=\frac{1}{T} \log \abs{C(\rho)}$. 
\end{definition}

Paralleling the pre-log factor $\min\{n_t,n_r\}$ in \eqref{eq:capacity}, which is commonly known as the total {\em number of degree of freedom} \cite{ZT}, we say the coding scheme $\{C(\rho)\}$ achieves the DMT of {spatial multiplexing gain} $r$ and \emph{diversity gain} $d(r)$ if the rate satisfies
\[
\lim_{\rho \to \infty} \frac{R(\rho)}{\log(\rho)} = r,
\]
and the average error probability is such that
\[
P_e(\rho) \ \doteq \ \rho^{-d(r)},
\]
where by the dotted equality we mean $f(M) \doteq g(M)$ if 
\beq
\lim_{M\to \infty}\frac{\log(f(M))}{\log(M)} = \lim_{M\to \infty}\frac{\log(g(M))}{\log(M)}. \label{eq:dotdefn}
\eeq
Notations such as $\dot\geq$ and $\dot\leq$ are defined in a similar way. 
\begin{remark}
We will still use,  for example, $f(M) \dot\geq  g(M)$ even when the limit at the RHS of \eqref{eq:dotdefn} does not exist. By this we only mean that $g(M)$ can be upper bounded by some function $c(M)$ where $c(M) \doteq f(M)$.
\end{remark}

With the above, the most important result in \cite{ZT} is the following.
\begin{thm}[DMT \cite{ZT}] \label{thm:DMT}
Let $n_t$, $n_r$, $T$, $\{C(\rho)\}$, and $d(r)$ be defined as before. Then any STBC coding scheme $\{C(\rho)\}$ has error probability lower bounded by
\beq
P_e(r) \ \dot\geq\ \rho^{-d^*(r)} \label{eq:DMT1}
\eeq
or equivalently, the diversity gain
\beq
d(r) \leq d^*(r), \label{eq:DMT2}
\eeq
when the coding is limited within a block of $T \geq n_t + n_r-1$ channel uses. The function of the optimal diversity gain $d^*(r)$, also termed the optimal DMT, is a piece-wise linear function connecting the points $(r,(n_t-r)(n_r-r))$ for $r=0,1,\ldots,\min\{n_t,n_r\}$. 
\end{thm}

\begin{figure}[!h]
\[\includegraphics[width=0.8\columnwidth]{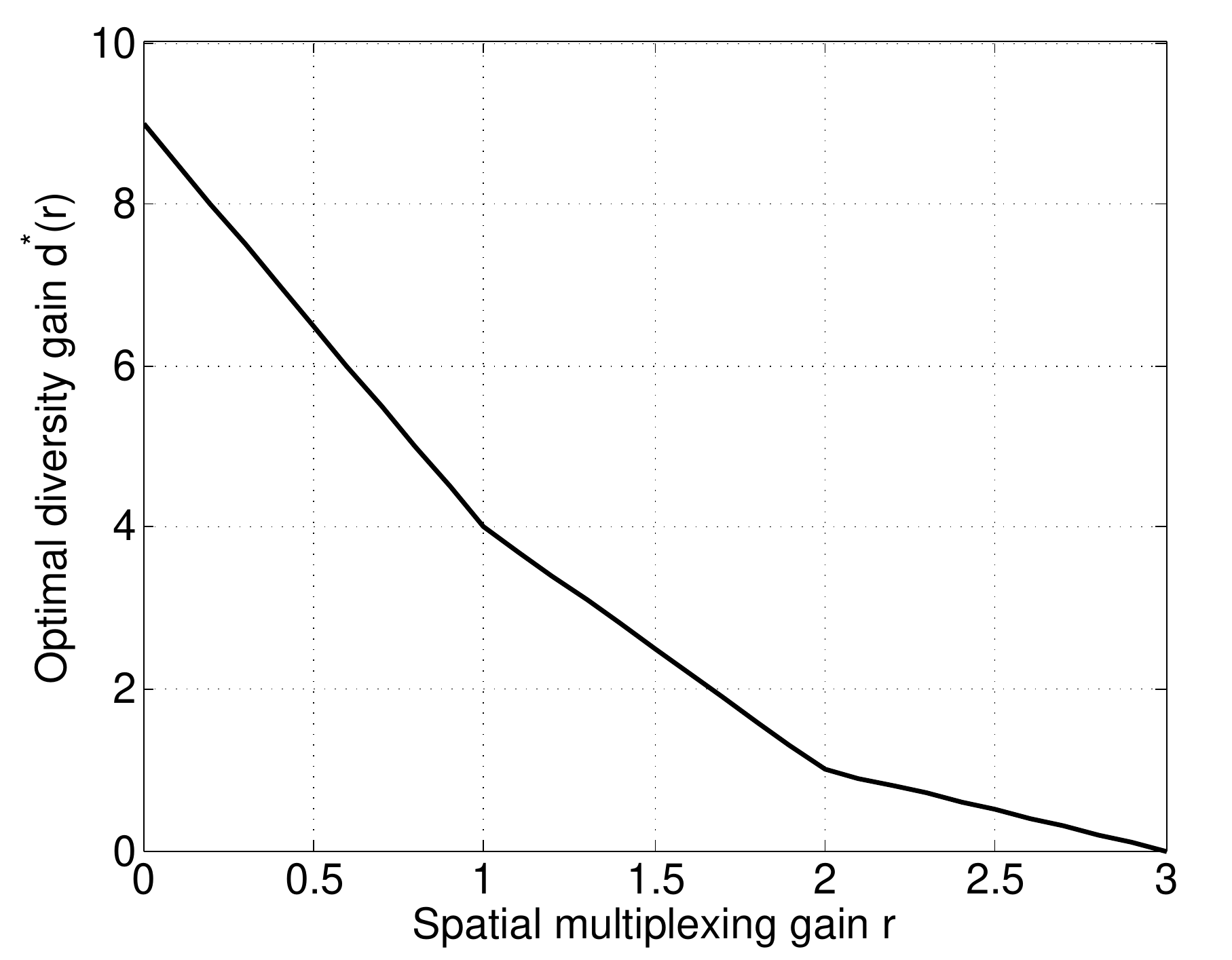}\]
\caption{DMT $d^*(r)$ for $n_t=n_r=3$.} \label{fig:DMT}
\end{figure}

An example of optimal DMT $d^*(r)$ for $n_t=n_r=3$ is given in Fig. \ref{fig:DMT}. We also remark that there exist space-time lattice codes \cite{EKPKL, BORV} that are optimal in the DMT sense, i.e., achieve the optimal diversity gain $d^*(r)$. The condition of $T$ in Theorem \ref{thm:DMT} has been improved to $T \geq n_t$ by Elia {\it et al. } in \cite{EKPKL}. Due to the outstanding error performance of space-time lattices codes, we shall study these codes in general in the next section. 

Before concluding this section, we make the following remark to further motivate the remainder of this paper. First, while the notion of DMT provides an asymptotic measure of the error performance of code $C(\rho)$ by focusing on the diversity exponent $d(r)$ as $\rho \to \infty$, there are certain limitations. For example, it is often observed in simulations that two coding schemes $\{C_1(\rho)\}$ and $\{C_2(\rho)\}$, having the same diversity gain $d(r)$, can differ significantly in error performance when SNR $\rho$ is finite. In other words, without conducting a simulation it is impossible to determine which code has better error performance at moderate SNR level from the DMT analysis. This happens especially when the error probability for $\{ C_1(\rho)\}$ takes the form of $P_{e_1}(r) = c_1(\rho) \rho^{-d(r)}$ and similarly $P_{e_2}(r) = c_2(\rho) \rho^{-d(r)}$ for $\{ C_2(\rho)\}$, and when the functions $c_1(\rho)$ and $c_2(\rho)$ behave like a constant in the asymptotic sense, i.e., in terms of the dotted noations
\[
c_1(\rho) \ \doteq \ c_2(\rho) \ \doteq 1. 
\]
On the other hand, the above asymptotic ambiguity can be resolved by the inverse determinant sum, which will be introduced in Section \ref{sec:3}.  Furthermore, it will be seen that the inverse determinant sum also represents an alternative, and probably better, criterion for designing STBC in general. 

\subsection{Matrix Lattices and spherically shaped coding schemes}\label{latticesection}

In this paper,  we will consider STBC with $n_t = T = n$, and therefore these codes live in the space $M_n(\C)$. 
With this choice, using results from classical lattice theory in $\R^{2n^2}$, we can define a natural inner product that induces the Frobenius norm in $M_n(\C)$.

We can \vv{flatten} $X\in M_n(\C)$ to obtain a $2n^2$-dimensional real
vector $\_{x}$ first by forming a vector of length $n^2$ out of the entries
(e.g. vectorizing row by row, or column by column) and then by replacing
each complex entry with the pair formed by its real and imaginary parts.
This defines a mapping $\alpha$ from $M_n(\C)$ to $\R^{2n^2}$:
\begin{equation}\label{eq:alpha}
\alpha: X \mapsto  \_{x} =\alpha(X)
\end{equation}
which is clearly $\R$-linear:
\begin{equation}\label{eq:Rlin}
\alpha(rX+r'X')=r\alpha(X)+r'\alpha(X'), \quad \forall r,r'\in\R.
\end{equation}
Let $\norm{X}_F = \sqrt{\tr(X^\dagger X)}$ denote the Frobenius norm of $X$.
Note that the following equality holds:
\begin{equation}\label{eq:alphaiso}
\norm{X}_F = \sqrt{\sum_{i=1}^n\sum_{j=1}^n|X_{ij}|^2}= \norm{\alpha(X)}_E,
\end{equation}
where $\norm{\cdot}_E$ denotes the Euclidean norm of a vector.
This makes $\alpha$ an isometry. It also gives us a natural inner product in the space $M_n(\C)$.
Given two matrices $X, Y \in M_n(\C)$, we define $\langle X, Y \rangle= \Re(\tr(XY^{\dagger}))=\langle \alpha(X), \alpha(Y)\rangle$, where the last notation $\langle \cdot \rangle$ stands for the natural Euclidean inner product in $\R^{2n^2}$.

\begin{definition}
A {\em matrix lattice} $L \subseteq M_n(\C)$ has the form
$$
L=\Z B_1\oplus \Z B_2\oplus \cdots \oplus \Z B_k,
$$
where the matrices $B_1,\dots, B_k$ are linearly independent over $\R$, i.e., form a lattice basis, and $k$ is
called the \emph{rank}  or the \emph{dimension} of the lattice.
\end{definition}

\begin{definition}\label{def:NVD}
If the minimum determinant of the lattice $L \subseteq M_n(\C)$ is non-zero, i.e. it satisfies
\[
\inf_{{\bf 0} \neq X \in L} \abs{\det (X)} > 0, 
\]
we say that the lattice satisfies the \emph{non-vanishing determinant} (NVD) property.
\end{definition}


We now consider a spherical shaping scheme based on a $k$-dimensional lattice $L$ inside $M_{n}(\C)$. 
Given a positive real number $M$ we define
$$
L(M)=\{a \in L \;:\; \norm{a}_F \leq M, a\neq {\bf 0} \}.
$$
We will also use the notation
$$
B(M)=\{a  \in M_n(\C) \;:\; \norm{a}_F \leq M \}
$$
for the sphere with radius $M$.

The following two results are well known.
\begin{lemma}[Spherical shaping]\label{spherical}
Let $L$ be a  $k$-dimensional lattice in  $M_{n}(\C)$ and
$L(M)$ be defined as above; 
then
\[
|L(M)|= cM^{k}+ O(M^{k-1}),
\]
where $c$ is some real constant, independent of $M$. 
\end{lemma}
\begin{IEEEproof}
For the proof we refer the reader to \cite{LP}.
\end{IEEEproof}

\begin{proposition}\label{Mclaurin}
Let $L$ be a $k$-dimensional lattice in $M_n(\C)$. Then
\begin{eqnarray*}
H_1 M^{s+k} \leq \sum_{X\in L(M)} \norm{X}_F^s \leq H_2M^{s+k}, \,\, s+k>0 \\
H_3 \log(M) \leq \sum_{X\in L(M)} \norm{X}_F^{s} \leq H_4 \log(M), \,\, s+k=0 \\
\sum_{X\in L(M)} \norm{X}_F^s \leq H_5, \,\, s+k<0, 
\end{eqnarray*}
where $H_i$ are constants independent of $M$.
\end{proposition}
\begin{IEEEproof}
The proof is a basic exercise in lattice theory. We refer the reader to \cite{LP} for the needed background.
\end{IEEEproof}

In particular, it follows that we can choose real constants $K_1$ and $K_2$ such that
\[
K_1M^k\geq |L(M)|\geq K_2 M^k.
\]

For subsequent discussions, the following definition will be useful. 
\begin{definition}
Suppose that $L$ is a $k$-dimensional lattice in $\R^n$. The function $f:\C\to \C$, where
$$
f(s)=\sum_{\_{x} \in L, \_{x}\neq \_{0}} \norm{\_{x}}_E^s,
$$
is well defined, when $-\Re(s)>k$ and is called the \emph{Epstein zeta function} \cite{Epstein}.
\end{definition}

With the above, we are now prepared to give a formal definition of a family of space-time lattice codes of finite size.
\begin{definition}
Given the lattice $L \subset M_n(\C)$, a space-time lattice coding scheme associated with $L$ is a collection of STBCs where each member is given by
\begin{equation}\label{codingscheme}
C_L( \rho)=\rho^{-\frac{rn}{k}}L\left(\rho^{\frac{rn}{k}}\right)
\end{equation}
for the desired multiplexing gain $r$ and for each $\rho$ level.
\end{definition}

The normalization factor $\rho^{-\frac{rn}{k}}$ in \eqref{codingscheme} is ony appropriate, but not exact, for meeting the average power constraint \eqref{powernorm}. Specifically, one might wonder whether the STBC $C_L( \rho)$ has average power exceeding the upper constraint in \eqref{powernorm} or it can still be improved. From  Proposition \ref{Mclaurin} we have
$$
 \sum_{X \in L\left(\rho^{\frac{rn}{k}}\right)} \rho^{-\frac{2rn}{k}} \norm{X}_F^2 \doteq \rho^{-\frac{2rn}{k}}(\rho^{rn/k})^{k+2}=\rho^{rn}.
$$
On the other hand we also have that $|L(\rho^{\frac{rn}{k}})| \doteq\rho^{rn}$ from Proposition \ref{spherical}. Combining the above shows that the code  $C_L( \rho)$ has the correct average power from the DMT perspective, i.e., in terms of the dotted equality. Henceforth, we simply ignore the scaling factor $\frac{1}{n_t}$ of $\SNR$ in the channel equation \eqref{eq:channel} as it is irrelevant to DMT calculations.

\section{ Inverse Determinant Sums Over Matrix Lattices} \label{sec:3}
In this section we introduce inverse determinant sums, study their basic properties and show how they  are related to DMT. We first begin with a non-rigorous introduction, which shows how these sums  appear naturally  as a continuation of more familiar sums.

Consider a $k$-dimensional lattice code $L(M)\subset \C^n$ for 
the following additive complex Gaussian noise channel
\[
\_{y} \ = \ \_{x} + \_{n}
\]
where $\_{x} \in L(M)$ and $\_{n}$ is a length-$n$ complex Gaussian random vector with zero mean and covariance matrix $I_n$. 

We have the familiar expression of the pairwise-wise error probability (PEP) upper bound for confusing $\_{x}$ to $\_{x}'$ at the receiver
\[
P( \_{x} \to \_{x}')\leq e^{-\norm{\_{x}-\_{x}'}_E^2}.
\]
If the codewords from the code $L(M)$  are sent equiprobably, we can upper bound the average error probability  by the following sum
\[
P_e \leq \sum_{\_{x} \in L,\, 0 < \norm{\_{x}}_E \leq 2M} e^{-\norm{\_{x}}^2},
\]
where the  term $2M$ follows as we have to consider differences of codewords. The right-hand-side is indeed a well known truncated \emph{exponential sum} taking values on lattice points.

\medskip
The second example channel is a  quasi static Rayleigh fading channel with single transmit and $n_r$ receive antennas. Assume that the channel vector is known perfectly to the receiver but not to the transmitter. We then have for the code $L(M) \in \C^n$
\[
P( \_{x} \to \_{x}')\leq \frac{1}{\norm{\_{x}-\_{x}}_E^{2n_r}},
\]
and the corresponding upper bound on overall error probability
\[
P_e \leq \sum_{ \_{x} \in L,\, 0 < \norm{\_{x}}_E \leq 2 M}\frac{1}{\norm{\_{x}}_E^{2n_r}}.
\]
We can then see that if $2n_r>k$,  the RHS is the truncated  Epstein zeta function. 
\medskip

We now turn to the more general case of having a $k$-dimensional NVD lattice  $L$ and consider finite code $L(M) \subset M_n(\C)$ and a slow Rayleigh fading MIMO channel with $n$ transmit and $n_r$  receive antennas. The channel equation can then be written as
\[
Y \ = \ H X + N,
\]
where $H$ and $N$ are respectively the channel and noise matrices  and where $X \in L(M)$. In terms of PEP, we have for $X \neq X'$
\[
P(X \to X')\leq \frac{1}{| \det(X-X')|^{2n_r}}, 
\]
and the corresponding upper bound on overall error probability 
\[
P_e \ \leq \ \sum_{X\in L,\, 0 < \norm{X}_F\leq 2M}\frac{1}{ |\det(X)|^{2n_r}}.
\]
We summarize the three cases above below. 
\begin{itemize}
\item Single antenna channel AWGN: $P_e$ is upper bounded by the sum of $e^{-\norm{\_{x}}_E^{2}}$, an \emph{exponential sum}.
\item Single antenna slow fading channel: $P_e$ is upper bounded by the sum of   $\frac{1}{\norm{\_{x}}_E^{2n_r}}$, an \emph{Epstein  zeta function}.
\item Quasi-static Rayleigh fading MIMO channel: $P_e$ is upper bounded by the sum of $\frac{1}{|\det(X)|^{2n_r}}$, an \emph{inverse determinant sum}.
\end{itemize}
We will see that the behavior of the third sum is the most peculiar. While in the second case the limit of the sum for $M \to \infty$ can be made to converge by increasing $n_r$, in the last case of inverse determinant sums we will show  that they might not converge.

\subsection{ The Basic Problem}\label{basicprob}

Let $L\subseteq M_n(\C)$ be a $k$-dimensional lattice. For any fixed $m\in \Z^+$ we define 
$$
S_L^m(M):=\sum_{X\in L(M)} \frac{1}{|\det(X)|^{m}}.
$$
Our main goal is to study the growth of this sum as $M$ increases. 
In particular, we are interested to find, if possible, a function $f(M)$ such that
$$
 S_L^m(M) \doteq f(M).
$$
As we will later see this ``dotted'' accuracy is  enough to determine the DMT of the code under consideration. Furthermore, it gives us a way to select codes with better error performance. Suppose that we have two  $k$-dimensional lattices $L_1$ and $L_2$, and corresponding functions $ S_{L_1}^m(M) \doteq f_1(M)$  and $ S_{L_2}^m(M) \doteq f_2(M)$. It is not far fetched to assume that if $f_1(M) \dot> f_2(M)$, the lattice $L_2$ would be a better code, at least for large code sizes.

Let us, however, shortly  discuss  inverse determinant sums in  a more accurate sense. Let us denote with $\Vol(L)$ \emph{the volume of the fundamental parallelotope} of a $k$-dimensional lattice $L$ in $M_n(\C)$. The normalized version of the inverse determinant sums problem is then to consider  the growth of the sum
\begin{equation}\label{volform}
\tilde{S}_L^m(M)= \Vol(L)^{mn/k}\sum_{X\in L(M)} \frac{1}{|\det(X)|^{m}}.
\end{equation}
Here the relevant accuracy level is to find, if possible, functions $f(M)$ and $g(M)$, where $ \mathop{\lim }\limits_{M \to \infty }g(M)/f(M)=0$, such that
$$
 | \tilde{S}_L^m(M)-f(M)|\leq g(M).
$$
Again it is reasonable to surmise that the smaller the function $f(M)$, the better the corresponding code will be. Comparing codes in this sense does take into account the size of the normalized minimum determinant and the number of times this worst case appears. Obviously comparing two codes in this normalized sense is more reliable than  comparing two codes in the previously described dotted sense. However, only in Section \ref{constremark} we will consider inverse determinant sums with an accuracy needed for this analysis.

\subsection{An  example of the effect of the difference in the growth of inverse determinant sums on the performance of space-time codes}\label{example}

The work in this paper is mostly theoretical,
but let us give an example that suggests that the inverse determinant sum is also a very practical research subject. 

Consider the following lattices
\begin{align*}
&L_1=\left\{\begin{pmatrix}
x_1&3x_2^*\\
x_2& x_1^*
\end{pmatrix}:\quad x_1,x_2 \in \Z[i]\right\},\\
&L_2=\left\{\begin{pmatrix}
x_1&-3x_2^*\\
x_2& x_1^*
\end{pmatrix}:\quad x_1,x_2 \in \Z[i]\right\}.
\end{align*}
Both $L_1$ and $L_2$ are $4$-dimensional lattice codes  in $M_2(\C)$, and as lattices they are isometric and have exactly the same normalized minimum determinant. Suppose that these codes are to be used for communication on a Rayleigh fading channel with a single receive antenna.
The corresponding inverse determinant sums  are of the type
$$
\sum_{X\in L_i(M)} \frac{1}{|\det(X)|^{2}}.
$$
We will later see that 
$$
S_{L_1}^2(M) \doteq M^2 \,\,\textnormal{ and }\,\,S_{L_2}^2(M)\doteq M^0.
$$
Here from the normalized minimum determinant  and shaping point of view these two codes are identical. Yet,  their inverse determinant sums differ dramatically and suggest that the code $C_{L_2}$ derived from the lattice $L_2$ has error performance much better than $C_{L_1}$ derived from $L_1$. The question is whether this difference will be visible in practice. After all, these inverse determinant sum considerations have an asymptotic nature.

\begin{figure}[!h]
\includegraphics[width=\columnwidth]{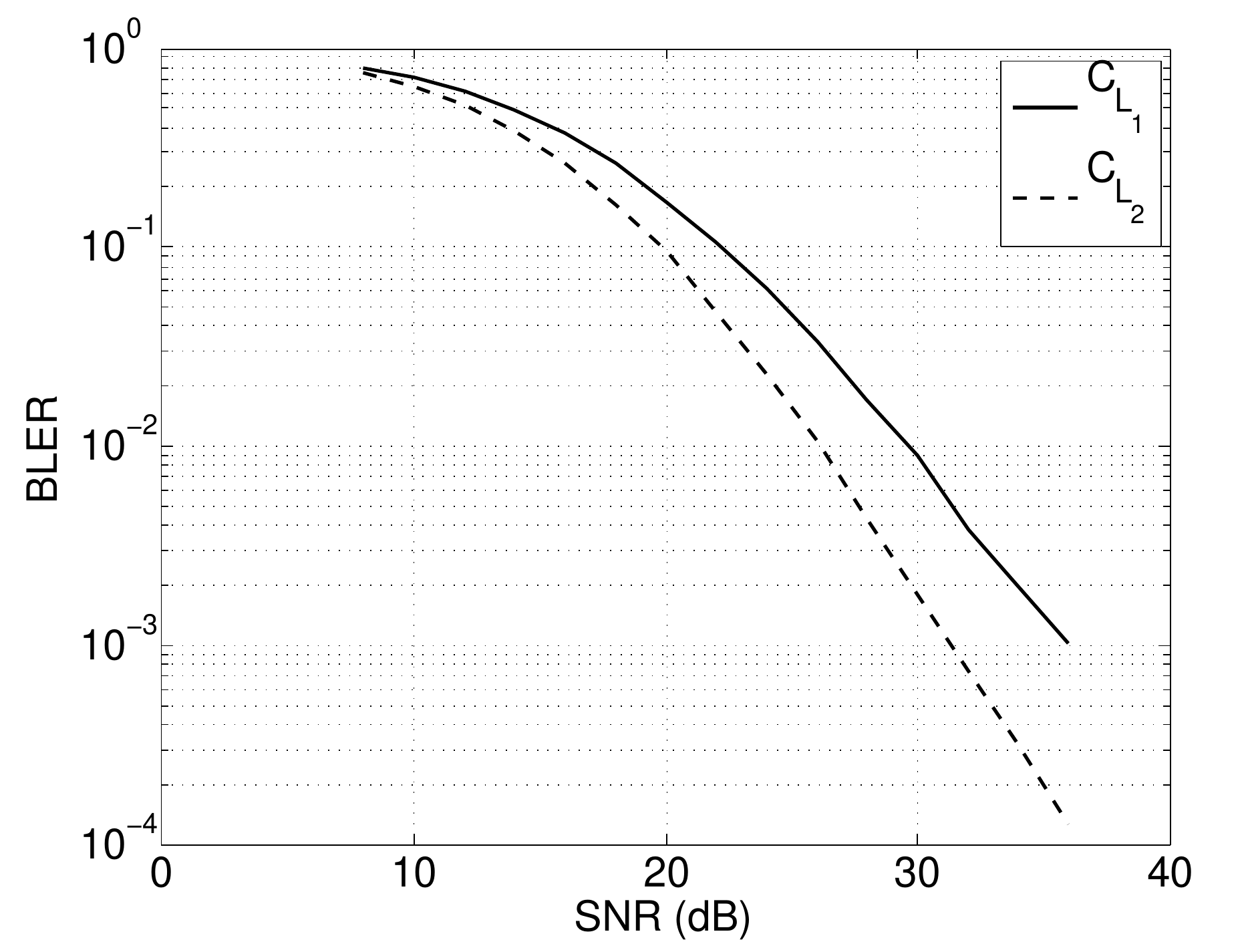}
\caption{Block error rates of codes $C_{L_1}$  and $C_{L_2}$ at 4 bpcu.}
\label{simulakuva}
\end{figure}

In Fig. \ref{simulakuva} we see the performance  of $C_{L_1}$ and $C_{L_2}$ where the components $x_1$ and $x_2$ takes values from the 16-QAM modulation. It can be clearly seen that $C_{L_2}$ performs much better than $C_{L_1}$ as predicted by the inverse determinant sums. 

\subsection{Elementary  bounds and some basic results for inverse determinant sums}

We now provide some simple upper and lower bounds for the  asymptotic behavior of $S_L^m(M)$ for a $k$-dimensional NVD-lattice $L$ in $M_n(\C)$.

\begin{proposition}\label{upperlower}
Suppose that $L$ is a $k$-dimensional  NVD-lattice  in $M_n(\C)$, with 
\[
\text{mindet}(L) \ := \ \min_{{\bf 0} \neq X \in L} \abs{\det(X)} \ = \ 1.
\]
We then have that 
\begin{eqnarray*}
K M^k \geq \sum_{X\in L(M)} \frac{1}{|\det(X)|^{m}}\geq K_1 M^{k-mn}, \,\, k-mn>0 \\
K M^k \geq \sum_{X\in L(M)} \frac{1}{|\det(X)|^{m}}\geq K_2 \log(M), \,\, k-mn=0\\
K M^k \geq \sum_{X\in L(M)} \frac{1}{|\det(X)|^{m}}\geq K_3, \,\,\,\,  k-mn<0,
\end{eqnarray*}
for some constants $K$, $K_1$, $K_2$, and $K_3$.

\end{proposition}
\begin{IEEEproof}
Hadamard inequality combined with the arithmetic mean-geometric mean (AM-GM) inequality gives us
$$
|\det(X)|\leq \left(\frac{\norm{X}_F}{\sqrt{n}}\right)^n.
$$

We then have that 
$$
\sum_{X\in L(M)} \frac{1}{|\det(X)|^{m}}  \geq \sum_{X\in L(M)} \frac{\sqrt{n}^{mn}}{\norm{X}_F^{nm}}.
$$
Applying Proposition \ref{Mclaurin} yields the lower bounds.

On the other hand, if $|\det(X)|=1$ for all nonzero $X\in L$ as the worst case, then 
$$
\sum_{X\in L(M)} \frac{1}{|\det(X)|^{m}}=\sum_{X\in L(M)} 1 =|L(M)|\leq K M^k,
$$
where $K$ is a constant independent of $M$ and where the last inequality follows from Lemma \ref{spherical}.
\end{IEEEproof}

We next provide an unsurprising invariance result, revealing that 
the growth of the inverse determinant sum of a matrix lattice $L\subset M_n(\C)$ is similar to the corresponding growth of the lattice $AL$, where $A$ is an invertible matrix in $M_n(\C)$.
We need a few lemmas.

\begin{lemma}\cite{KW}\label{mismatch}
Let $A$ and $B$ be invertible  matrices in  $M_n(\C)$ and let
$a_1\geq\dots\geq a_n$ be the eigenvalues of $AA^{\dagger}$ and $b_1\leq\dots \leq b_n$ be the eigenvalues of $BB^{\dagger}$.
We then have that
\[
\norm{AB}_F^2\geq \sum_{i=1}^{n} a_i b_i.
\]
\end{lemma}

\begin{lemma}\label{pulla}
Suppose that $\cal X$ is a set of matrices in $M_n(\C)$ and that $A$ is an invertible matrix in $M_n(\C)$. If
$f$ is a function such that for all $M > 0$ 
\[
\abs{B(M)\cap {\cal X}} \leq f(M), 
\]
then there exists a constant $K$ such that for all $M$
\[
\abs{B(M)\cap A{\cal X}} \leq \, f(KM),
\]
where $A {\cal X} = \{ AX : X \in {\cal X}\}$.
\end{lemma}
\begin{IEEEproof}
Let $\lambda_1$ be the smallest eigenvalue of $AA^{\dagger}$. Lemma \ref{mismatch} implies that
for all the elements   $AX \in A{\cal X}$,  $\norm{AX}_F^2 \geq \lambda_1 \norm{X}_F^2 $. It follows that
for the matrix  $AX$, where 
$$
\norm{AX}_F \leq M,
$$
we must have that $\norm{X}\leq\frac{M}{\sqrt{\lambda_1}}$. We now see  that $\frac{1}{\sqrt{\lambda_1}}$ is a suitable constant for $K$.
\end{IEEEproof}

\begin{proposition}\label{Alattice}
Let $L\subset M_n(\C)$ be a matrix lattice and $A \in M_n(\C)$ be an invertible matrix. If  $S_L^m(M) \doteq M^k$ for some $k$, then
$$
 S_{AL}^m(M) \stackrel{.}{=} M^k.
$$
\end{proposition}
\begin{IEEEproof}
Let $\lambda_1$ be the smallest eigenvalue of $AA^{\dagger}$. Using the same argument as in the previous lemma we have
$$
\sum_{X\in AL(M)} \frac{1}{|\det(X)|^{m}}\leq \sum_{Y\in L(M/\sqrt{\lambda_1})} \frac{|\det(A)|^{-m}}{|\det(Y)|^{m}}.
$$
Changing  the roles of $AL$ and $L$ and replacing $A$ with $A^{-1}$ give the other direction of the inequality
\begin{align}
S_L^m(M)  \dot\leq S_{AL}^m(M). \tag*{\IEEEQED}
\end{align}
\let\IEEEQED\relax
\end{IEEEproof}
The previous proposition obviously works also in the case where the lattice $L$ is multiplied by a matrix $A$ from the right.

\subsection{Inverse determinant sum in relation to DMT} \label{subsec:3}
In this section we will show how  we can use  DMT to prove lower bounds for the asymptotic growth of inverse determinant sums. At the same time we will get a criterion for a code to achieve the optimal  DMT for multiplexing gains $r\in [0,1].$

Let  $L\subseteq M_n(\C)$ be a $k$-dimensional lattice, and consider the finite codes $C_L(\rho)$
defined in (\ref{codingscheme}). 
Assume there are $n_r$ receive antennas. Then following the union bound together with the PEP based determinant inequality \cite{TSC}, we get the following bound for the average error probability for the code $C_L(\rho)$
\begin{equation} \label{union_bound}
P_e \leq \rho^{-nn_r(1-2nr/k)}\sum_{X\in L(2\rho^{rn/k})} \frac{1}{|\det(X)|^{2n_r}}.
\end{equation}

The moral of the following proposition is that the determinant sum of  a space-time lattice code must grow with considerable speed,  or otherwise the code  would have DMT exceeding $d^*(r)$ given in Theorem \ref{thm:DMT}. 
 \begin{proposition}\label{mainDMT}
Let $L$ be a $k$-dimensional  fully diverse lattice in $M_n(\C)$ and $n_r$ be a positive integer.  Suppose that  $S_L^{2n_r}(M)\doteq M^v$ for some $v\in \R$. We then have that
\[
S_L^{2n_r}(M)=\sum_{X\in L(M)} \frac{1}{|\det(X)|^{2n_r}}  \dot\geq \ M^{(n_rk/n+k -k/n - 2nn_r)}.
 \]
\end{proposition}
\begin{IEEEproof}
Consider the previously mentioned coding scheme $C_L( \rho)=\rho^{-\frac{rn}{k}}L\left(\rho^{\frac{rn}{k}}\right)$. As we have shown, the union bound \eqref{union_bound} yields the following lower bound for $S_L^{2n_r}(M)$
\[
S_L^{2n_r}(M) \ \geq \ P_e \ \cdot \ \rho^{nn_r(1-2nr/k)} 
\]
where $M=2\rho^{nr/k}$. Theorem \ref{thm:DMT}, on the other hand, shows that for integer values of $r$ 
$$
P_e \dot\geq \rho^{-(n-r)(n_r-r)}.
$$
Combining the above gives for integer values of $r$.
\begin{align*}
& S_L^{2n_r}(2\rho^{nr/k}) \stackrel{.}{\geq}  \rho^{-((n-r)(n_r-r)- nn_r(1-2nr/k))  }\\
&= \rho^{-(r^2-nr -rn_r +2n^2rn_r/k)}.
\end{align*}
Hence,
 $$
 S_L^{2n_r}(M)\stackrel{.}{\geq}  M^{-(rk/n-k-n_rk/n+ 2nn_r)}.
 $$
 The maximum here is achieved obviously for $r=0$, but in this case we do not have growth for our matrix sum as the corresponding $M=1$. The next integer point is  $r=1$. In this case we have 
 \begin{align}
 S_L^{2n_r}(M)\ \dot \geq\ M^{(n_rk/n+k -k/n - 2nn_r)}. \tag*{\IEEEQED}
 \end{align}
 \let\IEEEQED\relax 
 \end{IEEEproof}

\begin{corollary}\label{DMTcorollary}
Let $L \subset M_n(\C)$ be a $k$-dimensional fully diverse lattice. If the corresponding inverse determinant sum achieves the lower bound in Proposition \ref{mainDMT}, then $C_L(\rho)$ achieves the optimal DMT for $r\in [0,1]$, when received with $n_r$ antennas. 
\end{corollary}
\begin{IEEEproof}
Here we have  $S_L^{2n_r}(M)\dot = M^{(n_rk/n+k -k/n - 2nn_r)}$. Setting  $M=2\rho^{rn/k}$ and substituting the above into \eqref{union_bound} yield
\bean
P_e & \dot \leq& \rho^{-nn_r(1-2nr/k)}\rho^{(rn/k)(n_rk/n+k -k/n - 2nn_r)}\\
&=& \rho^{-nn_r +r(n_r+n-1)}.
\eean
Comparing the above to the DMT lower bound, $P_e \geq \rho^{-d^*(r)}$ for $r \in [0,1]$, where $d^*(r)$ is a straight line connecting the points $(0,nn_r)$ and $(1, (n_r-1)(n-1))$ yields the desired result. 
\end{IEEEproof}

\begin{remark}
We have stated  Proposition \ref{mainDMT} in the simplest possible form by assuming $\frac{\log(S_L^{2n_r}(M))}{\log(M)}$ has a limit when $M$ approaches infinity. While this condition is not that restrictive, the proof of Proposition \ref{mainDMT} gives us more. It actually states that if there is a function $f(M)$, having a limit in the dotted sense, for which $S_L^{2n_r}(M)\leq f(M)$, then  $f(M)\stackrel{.}{\geq} M^{(n_rk/n+k -k/n - 2nn_r)}$. In particular we cannot upper bound $S_L^{2n_r}(M)$ with any $KM^t$, where $t<(n_rk/n+k -k/n - 2nn_r)$ and $K$ is some constant.

\end{remark}

\subsection{The inverse determinant sum and DMT of  the Alamouti code}

In this section we will show that the Alamouti code does reach the  bound in Proposition \ref{upperlower}. This result then allows us to rediscover the DMT of Alamouti code when received with $n_r$ antennas.

The $2 \times 2$ Alamouti code is the following
\[
A(x_1, x_2, x_3, x_4)=
\begin{pmatrix}
x_1+x_2i& -(x_3 +x_4i)^*\\
x_3+x_4i& (x_1+x_2i)^*
\end{pmatrix}.
\]
for some indeterminate $x_1$, $x_2$, $x_3$ and $x_4$, where $i = \sqrt{-1}$. 
The corresponding lattice of the Alamouti code can be written as 
$$
L_{Alam}=\Z A(1,0,0,0)+\cdots+ \Z A(0,0,1,0)+\Z A(0,0,0,1),
$$
which  is a $4$-dimensional lattice in $M_2(\C)$.
We then consider the corresponding inverse determinant sum
$$
\sum_{X\in L_{Alam}(M)}\frac{1}{|\det(X)|^{2m}}.
$$

\begin{proposition}\label{alamsum}
Let $m\geq 1$ be a real number. Then,
$$
K_2\leq \sum_{X\in L_{Alam}(M)}\frac{1}{|\det(X)|^{2m}}\leq K_1 \log(M),
$$
where $K_1$ and $K_2$ are  some constants.
\end{proposition}

\begin{IEEEproof}
Due to the orthogonality of the rows of the Alamouti code, for any codeword $ X \in L_{Alam}$ we have 
$$
|\det(X)|= \left(\frac{\norm{X}_F}{\sqrt{2}}\right)^2.
$$
We now have that 
$$
 \sum_{X\in L_{Alam}(M)}\frac{1}{|\det(X)|^{2m}}=\sum_{X\in L_{Alam}(M)}\frac{2^{2m}}{\norm{X}_F^{4m}}.
$$
The rest follows from  Proposition \ref{Mclaurin}.
\end{IEEEproof}

\begin{remark}
In particular if $m$ is large enough the inverse determinant sum of the Alamouti code  is the Epstein zeta function.
\end{remark}

\begin{corollary}\label{AlamDMT}
 When received with $n_r$  antennas, the Alamouti code achieves the  DMT curve
$$
(r,  2n_r(1-r)), \quad 0 \leq r \leq 1,
$$
which is optimal in DMT for any $4$-dimensional lattice codec in $M_2(\C)$.
\end{corollary} 

\begin{IEEEproof}
In order to study the DMT of codes derived from the lattice $L_{Alam}$, we consider the spherical coding scheme $L_{Alam}(\rho^{r/2})\rho^{-r/2}$.
The usual union bound argument \eqref{union_bound} then implies
$$
P_e \leq \sum_{ L_{Alam}(2\rho^{r/2})} \frac{\rho^{-2n_r(1-r)}}{|\det(X)|^{2n_r}}.
$$
Also, by Proposition \ref{alamsum} we have 
$$
P_e \leq \rho^{-2n_r(1-r)}K(\log(2\rho^{r/2})),
$$
where $K$ is some constant independent of $\rho$.  This gives us that the Alamouti code does achieve the claimed DMT. The rest follows from \cite[Proposition  3.3]{VLISIT} where it is shown that this is the best possible for all $4$-dimensional lattice codes in $M_2(\C)$.
\end{IEEEproof}

\subsection{Applying DMT to  lattice theory}\label{mathematicalcomment}
In this subsection we give an example showing how Proposition \ref{mainDMT} can be used in lattice theory. 

Let  $L \subset \C^4$ be an $8$-dimensional lattice and consider the function $f:\C^4 \to \R$ by 
$$
f(x_1, x_2, x_3,x_4)=|x_1x_2-x_3x_4|.
$$
Assume that  $f(x_1, x_2, x_3,x_4)\neq 0 $ for any non-zero element in $L$. The reader can immediately see that $L$ can be reformulated as 
a matrix lattice and $f$ is the absolute value of the determinant of $2 \times 2$ matrices. We now see what can be said about the asymptotic behavior of   the sum
$$
\sum_{X\in L(M)}  f(X)^{-4}.
$$
By Proposition \ref{upperlower} we have 
$$
K_1M^8 \geq \sum_{X\in L(M)}  f(X)^{-4}\geq K_2 \log(M),
$$
for some constants $K_1$ and $K_2$. 
However, if we can bound the growth  of   $\sum_{X\in L(M)} f(X)^{-4}$   with any $K M^t$, where $K$ and $t$ are constants, then Proposition \ref{mainDMT} tells us that $t\geq 4$.  This reveals that the lower bound obtained from Proposition \ref{mainDMT} is considerably stronger than the one from Proposition \ref{upperlower} and tells us something non-trivial about the asymptotic behavior of this sum.

However, it should be noted that Proposition \ref{mainDMT} only applies to the cases when $m$ in  the sum $S_L^m(M)$ is an even integer. 

We will see that the lower bound is also the best possible in the sense that there are 8-dimensional lattices in $\C^4$ for which 
$$
\sum_{X\in L(M)}  f(X)^{-4}\stackrel{.}{=} M^4.
$$

It is very likely that there are more direct methods that give this result, but it is intriguing that we can derive such lattice theoretic result  from  information theory.

\section{Inverse determinant sums of algebraic number fields and DMT of diagonal codes} \label{sec:4}

We now consider inverse determinant sums arising from algebraic number field codes \cite{Belfiore}. In particular we will show how the error probability of these codes is tied to the unit group and Dedekind zeta function of the corresponding algebraic number field. These connections allow us to give a better look at the behavior of these codes and to prove their DMT optimality. The proof of this case will 
give some insight into the case of codes arising from division algebras.

For simplicity let us consider a degree $n$ cyclic number field extension $K/\Q(i)$, where the Galois group is $\langle\sigma\rangle = \{ \sigma_1, \cdots, \sigma_n\}$,  and $\OO_K$ is a principal ideal domain (PID). We will comment more on these conditions in Section \ref{constremark}. 

We can define a
\emph{relative canonical embedding} of $K$ into $M_n(\C)$ by
$$
\psi(x)=\diag(\sigma_1(x),\dots, \sigma_n(x)),
$$
where $x$ is an element in $K$.
The ring of algebraic integers $\OO_K$ has a  $\Z$-basis $W=\{w_1,\dots ,w_{2n}\}$ and therefore
$$
\psi(\OO_K)=\psi(w_1)\Z+\cdots +\psi(w_{2n})\Z,
$$
is a $2n$-dimensional lattice of matrices in $M_n(\C)$. The main reason to use such a construction is that for each nonzero element $a\in \OO_K$, we have that $|\det(\psi(a))|\geq 1$. 

Let $L=\psi(\OO_K) \subset M_n(\C)$ be the $2n$-dimensional  number field lattice code and consider the coding scheme $C_L(\rho)$ in (\ref{codingscheme}).
Before measuring the DMT  for this type of codes, the following definition will be useful. Let $I_{\OO_K}$ be the set of nonzero ideals of the ring $\OO_K$. The Dedekind zeta function is 
\begin{equation}\label{Dedekind_zeta_function}
\zeta_{K}(s)=\sum_{\A\in I_{\OO_K}}\frac{1}{[\OO_K:\A]^s},
\end{equation}
where  $s$ is a complex number with  $\Re(s)>1$.

We give the following example for illustration. 

\begin{exam}\label{QAM-zeta}
The simplest example of the previous construction arises from the trivial extension $\Q(i)/\Q(i)$. The Galois group then consists simply of 
the identity element. We then have a lattice $L=\Z[i] \subset \C$, which is  a 2-dimensional lattice in $M_1(\C)$. Furthermore, let $L(M/2)$ be the finite code derived from $L$. When received by $n_r$ antennas, the error probability of $L(M/2)$ has a union bound \eqref{union_bound} containing the following sum
\[
\sum_{x\in L(M)}\frac{1}{|xx^*|^{n_r}}=\sum_{x\in L(M)}\frac{1}{\norm{x}_E^{2n_r}}.
\]
The above is actually the truncated Epstein zeta function and calls for the bound in Proposition \ref{upperlower}
However, we can look at this problem from another angle that can be easily generalized. Notice that for every element  $x \in \Z[i]$, we  have $N_{\Q(i)/\Q}(x)=|x|^2$, hence
$$
\sum_{x\in L(M)}\frac{1}{\abs{x}^{2n_r}}=\sum_{x\in L(M)}\frac{1}{|N_{\Q(i)/\Q}(x)|^{n_r}}
$$
We know that $\Z[i]$ has only $4$ invertible elements $1,-1,i,-i$ and that $\Z[i]$ is a PID. Therefore, for every ideal 
$x\Z[i]$, we have exactly $4$ different generators $x, -x,ix$, and  $-ix$.
We can then write
$$
\sum_{x\in L(M)}\frac{1}{\norm{x}^{2n_r}}=\sum_{N(I)\leq M^2 }\frac{4}{N(I)^{ n_r}},
$$
which is related to the truncated Dedekind zeta function $\zeta_{\Q(i)}(s)$ at the point $s=n_r$. In particular when we let $M$ grow to infinity we get
that the  sum  $\sum_{x\in L(M)}\frac{1}{\abs{x}^{2n_r}}$ approaches   $4\zeta_{\Q(i)}(n_r)$. 
\end{exam}

We point out that this approach was earlier taken in 
\cite{RVCC}. Yet, it only applies to the case when the extension has degree 1. We will next show how this can be extended to more general number fields.

Consider a cyclic  extension $K/\Q(i)$, where $[K:\Q(i)]=n$. With $L=\psi(\OO_k)$ defined as before, let $L(M/2)$ be the finite code derived from $L$. When received by $n_r$ antennas, the error probability of $L(M/2)$ has a union bound \eqref{union_bound} containing the following sum
\bea
&& \sum_{\norm{\psi(x)}_F\leq M,\; x\in \OO_K} \frac{1}{|\det(\psi(x))|^{2n_r}}\nonumber\\
& = & \sum_{\norm{\psi(x)}_F\leq M,\; x\in \OO_K} \frac{1}{|N_{K/\Q}(x)|^{n_r}}\nonumber\\
& =& \sum_{x\in X(M)}\frac{A_x(M)}{|N_{K/\Q}(x)|^{n_r}}, \label{eq:tbc}
\eea
where $X(M)$ is a  set of elements  $x\in \OO_K$,  $\norm{\psi(x)}_F\leq M$ , each generating a separate ideal in  $\OO_K$. Accordingly, $A_x(M)$ is the number of elements $y\in \OO_K$,  $\norm{\psi(y)}_F\leq M$, each generating the same principal ideal as the one generated by $x$.

If we neglect for the moment the terms $A_x(M)$, and consider only the sum $\sum_{x\in X(M)}\frac{1}{|N_{K/\Q}(x)|^{n_r}}$, we see that it is a part of the Dedekind zeta function $\zeta_{K}$ at the point $n_r$.

In the following we will give bounds for $A_x(M)$ and the truncated sum. The bounds depend only on the value of $M$; they are independent of the choice of $x$.

\subsection{Bounds  for $A_x(M)$ and truncated Dedekind zeta function}\label{orbit}

We begin our analysis with $A_1(M)$. This is the number of elements $u$ in the \emph{unit group} $\OO_K^*$ of the ring $\OO_K$ 
such that $\psi(u) \in B(M)\subset M_n(\C)$.

\begin{lemma}\label{units}

Let $K/\Q(i)$ be a cyclic field extension with $[K:\Q(i)]=n$. Then the cardinality of the set
\[
{\cal A}_1(M)=\{\psi(u) :  u\in \OO_K^*, \, \norm{\psi(u)}_F\leq M\}
\]
has an upper bound
\[
A_1(M)\ = \ \abs{{\cal A}_1(M)} \ \leq\  N \left( \log (M) \right)^{n-1},
\]
where $N$ is a constant independent of $M$.
\end{lemma}
\begin{IEEEproof}
For ease of reading, the proof to this lemma is relegated to Section \ref{sec:proof}. 
\end{IEEEproof}

Based on  Lemma \ref{units}, we can upper bound the value of $A_x(M)$ for all $x$. 

\begin{proposition}\label{Ux}
Let $K/\Q(i)$ be a cyclic field extension with $[K:\Q(i)]=n$ and let $x\in \OO_K$ be a non-zero element with $\norm{\psi(x)}_F\leq M$. Then 
\begin{eqnarray*}
A_x(M)  &=& |\{u : \norm{\psi(xu)}_F \leq M, u \in \OO_K^* \}| \\
& \leq&  N'\left( \log M \right)^{n-1},
\end{eqnarray*}
where $N'$ is a constant independent of $M$ as well as of  the element $x$.
\end{proposition}
\begin{IEEEproof}
Given $x \in \OO_K$, we can write  $\psi(x)=\diag(x_1,\dots, x_n)$.
The condition $\norm{\psi(x)}_F\leq M$ implies $|x_i|\leq M$ for all $i$. We also have that 
$ |x_1|\cdots |x_n|\geq 1$. It follows that for all $i$
\begin{equation}\label{coordinatesize}
 |x_i|\geq \frac{1}{M^{n-1}}.
 \end{equation}
Now let $u \in \OO_K^*$ be a unit such that
$\norm{\psi(ux)}_F=\norm{\psi(u)\psi(x)}_F=\norm{\diag(x_1 u_1,\dots, x_nu_n)}_F\leq M$, where $u_i = \sigma_i(u)$. 
We have that $\abs{x_i} \abs{u_i} \leq M$ for all $i$, and \eqref{coordinatesize} implies $ |u_i|  \leq M^n$. Therefore we have  that $\norm{\psi(u)}_F \leq \sqrt{n}M^n$. Lemma \ref{units} now gives that  
\bean
A_x(M) & \leq & A_1 \left(  \sqrt{n}M^n \right)\\
& \leq & N \left(\log(\sqrt{n}M^n)\right)^{n -1}\\
& \leq & N' \left( \log M \right)^{n-1},
\eean
where $N'$ is a constant independent of $M$.
\end{IEEEproof}

The essential part of this result is that we could find a constant $K$ such that  $K\left( \log M\right)^{n-1}$ upper-bounds $A_x(M)$ for all $x\in \OO_K$ with $\norm{\psi(x)}_F\leq M$.

Let us now give a bound for the truncated Dedekind zeta function.

\begin{proposition}\label{zeta}
Let $K/\Q(i)$ be a cyclic field extension with $[K:\Q(i)]=n$.  Then
\beq
\sum_{x \in X(M)} \frac{1}{|\det (\psi(x))|^{2n_r}} \leq N \left( \log M \right)^{2n}, \label{upperzeta}
\eeq
where $X(M)$ is a subset of $\OO_K$ in which each element $x$ generates a separate integral ideal and satisfies $\norm{\psi(x)}_F \leq M$, as defined in \eqref{eq:tbc}, and $N$ is a constant independent of $M$.
\end{proposition}
\begin{IEEEproof}
The proof is relegated to Section \ref{sec:proof} for ease of reading. 
\end{IEEEproof}
We remark that  if $n_r>1$ the upper bound in \eqref{upperzeta} is trivial as the resulting Dedekind zeta function converges to a constant, and we can limit the truncated sum that that constant. See Subsection \ref{constremark} for a discussion.

\subsection{The inverse determinant sum and DMT of algebraic number field codes}

Armed with Proposition \ref{Ux} and Proposition \ref{zeta}, we are now ready to continue the derivation of \eqref{eq:tbc} to obtain an upper bound for the inverse determinant sum of number field codes.

\begin{proposition}\label{numberfieldsum}
Let $K/\Q(i)$ be a cyclic field extension with $[K:\Q(i)]=n$. Then
\[
\sum_{\norm{\psi(x)}_F\leq M,\, x\in \OO_K }\frac{1}{|\det(\psi(x))|^{2n_r}}\leq N \left( \log M \right)^{3n-1},
\]
where $N$ is some constant independent of $M$. 
\end{proposition}
\begin{IEEEproof}
Continuing from \eqref{eq:tbc} we have 
\bean
&& \sum_{\norm{\psi(x)}_F\leq M,\, x\in \OO_K} \frac{1}{|\det(\psi(x))|^{2n_r}}\\
& =& \sum_{x\in X(M)}\frac{A_x(M)}{|N_{K/\Q}(x)|^{n_r}}\\
& \leq & N_1 \left( \log M \right)^{n-1} \sum_{x\in X(M)}\frac{1}{|N_{K/\Q}(x)|^{n_r}}\\
& \leq & N_1 \left( \log M \right)^{n-1} N_2 \left( \log M \right)^{2n},
\eean
where the first inequality follows from Proposition \ref{Ux} to upper-bound $A_x(M)$ with a constant $N_1$, and the second inequality is due to Proposition \ref{zeta} with another constant $N_2$. 
\end{IEEEproof}

\begin{remark}
Here we have an example of $2n$-dimensional lattices where the growth of inverse determinant sums is logarithmic. Comparing this bound to  that in Proposition \ref{upperlower} we can see that we are somewhat close to lower bounds  if $n_r=1$, but are far from them if $n_r>2$. This suggests that the bounds in Proposition \ref{upperlower} are not very tight.
\end{remark}

Finally, we are ready to determine the DMT curve achieved  by number field codes derived from lattice $L$, by which we mean the following. Let $K/\Q(i)$ be a cyclic field extension of degree $n$, and consider the $2n$-dimensional lattice 
\[
L \ = \ \left\{ \diag(\sigma_1(x), \cdots, \sigma_n(x)) \ : \ x \in \OO_K \right\}.
\]
Given SNR $\rho$ and multiplexing gain $r$, let 
\[
C_L(\rho)=\rho^{-\frac r2}L(\rho^{\frac r2})
\]
be the corresponding finite code obtained by the spherical coding scheme \eqref{codingscheme}.

\begin{thm}\label{numberfieldsDMT}
If the receiver has $n_r$  antennas, the code $C_L(\rho)$ achieves the following DMT curve
\[
(r, n n_r(1-r)^+),
\]
where $(a)^+=\max\{a,0\}$. 
\end{thm}
\begin{IEEEproof}
Note that $L$ is an NVD lattice. It can be easily shown that the maximal pair-wise error probability achieved by $C_L(\rho)$ is $\doteq \rho^{-nn_r(1-r)}$ \cite{VLISIT}, hence $P_e\ \dot\geq\  \rho^{-nn_r(1-r)}$. For the upper bound on $P_e$, the usual union bound argument gives
\begin{eqnarray*}
P_e & \leq & \sum_{X\in L(2\rho^{\frac r2})} \frac{\rho^{-n_r n(1-r)}}{|\det(X)|^{2n_r}} \\
& = & \sum_{\norm{\psi(x)}_F\leq 2\rho^{\frac r2},\, x\in \OO_K }\frac{\rho^{-n_r n(1-r)}}{|\det(\psi(x)|^{2n_r}}\\
& \dot\leq & \rho^{-n_r n(1-r)} \left( \log(2\rho^{\frac r2}) \right)^{3n-1}\\
& \doteq & \rho^{-n_r n(1-r)} ,
\end{eqnarray*}
where the last dotted inequality follows from Proposition \ref{numberfieldsum} after neglecting the constant factor. Combining the upper and lower bounds on $P_e$ proves the claim. 
\end{IEEEproof}

\subsection{ A remark on the constant values}\label{constremark}
 
In Proposition \ref{numberfieldsum} we showed the following result
\[
\sum_{\norm{\psi(x)}_F\leq M,\, x\in \OO_K }\frac{1}{|\det(\psi(x))|^{2n_r}}\leq N \left( \log M \right)^{3n-1}.
\]
For cyclic extensions, where $\OO_K$ is PID, we point out the assumption of being cyclic is not needed anywhere. This bound is also true in the case where $\OO_K$ is not a PID, but  it is only looser. 

The proof of this result was quite elementary and satisfactory for our purposes, and it can be easily tightened. Below we briefly discuss how our methods can give quite tight asymptotic bounds for number field codes, when the number of receiving antennas is greater than $1$.  

We note that the term  $(\log{M})^{2n}$ from Proposition \ref{zeta} can simply be  replaced with $\zeta_K(n_r)$ (see the proof of Proposition \ref{zeta}), and the function $\zeta_K(n_r)$ converges to some constant when $n_r > 1$. This  already reduces the bound in Proposition \ref{numberfieldsum} to  $N\zeta_K(n_r)(\log{M})^{n-1}$. We can say furthermore a few words about the constant $N$. 

The main theorem in \cite{EV} gives us the following asymptotic bound  
\begin{equation}\label{unitbound1}
 A_1(M) =\frac{\omega n^{n-1}(\log{M})^{n-1}}{R (n-1)!} + O((\log{M})^{n-2}),
\end{equation}
where $\omega$ is the number of roots of unity in $K$ and $R$ is the \emph{regulator}.
 
Let us now prove that  uniformly for all nonzero $x \in \OO_K$ we have
\[
A_x(M)\leq \frac{\omega n^{n-1}(\log{M})^{n-1}}{R (n-1)!} + O((\log{M})^{n-2}).
\]

Let  $x$ be a non-zero element in $\OO_K$. We can then write that
$$
|\psi(x\OO_K^*)\cap B(M)|= |ay\psi(\OO_K^*)\cap B(M)|\leq |y\psi(\OO_K^*)\cap B(M)|,
$$
where $y$ is a diagonal matrix with property $|det(y)|=1$ and $a\geq 1$ is a positive constant.  This means that  $f(y)$, where $f$  is extended to map from $\C^n$ (see  Section \ref{sec:proof} for a definition) is in the plane generated by some basis of 
$f(\OO_K^*)$.  The lattice $f(\OO_K^*)$ has a fixed covering radius $r$. Let us now suppose that $u$ is an element in $f(\OO_K^*)$ closest 
to $y$. This means that $||f(y)-f(u)||_E\leq r$. We can now write
$$
|y\psi(\OO_K^*) \cap B(M)|=  |y\psi(u^{-1})\psi(\OO_K^*) \cap B(M)|\leq
$$
$$ 
|\psi(\OO_K^*)\cap B(e^{nr} M)|\leq \frac{\omega (n)^{n-1}(\log{M})^{n-1}}{R (n-1)!} + O((\log{M})^{n-2}).
$$
This bound is dependent only on  $A_1(M)$ and $r$ and is therefore uniform for all $x\in \OO_K$.

Assume $n_r>1$. By collecting all the previous results we now have an upper bound
\begin{equation}\label{consteq}
S_{\psi(\OO_K)}^{2n_r}(M)\leq N \zeta_K(n_r) \left(\log{M}\right)^{n-1} + O((\log{M})^{n-2}) 
\end{equation}
 and a lower bound
$$
N  \left(\log{M}\right)^{n-1} + O((\log{M})^{n-2})\geq S_{\psi(\OO_K)}^{2n_r},
$$
where $N = \frac{ \omega (n)^{n-1}}{R (n-1)!}$. As  $Vol(\psi(\OO_K))=2^{-n}\sqrt{|d(K/\Q)|}$, equation \eqref{volform} gives us that
$$
\tilde{S}_{\psi(\OO_K)}^{2n_r}(M)\leq  N  \zeta_K(n_r) \left(\log{M}\right)^{n-1} + O((\log{M})^{n-2}) 
$$
and
$$
N\left(\log{M}\right)^{n-1} + O((\log{M})^{n-2})\geq \tilde{S}_{\psi(\OO_K)}^{2n_r},
$$
where $N = \frac{ \omega (n)^{n-1}}{R (n-1)!}(2^{-n}\sqrt{|d(K/\Q)|})^{n_r}$. 
 For the PID case this bound is probably quite tight asymptotically for the leading term $(\log(M))^{n-1}$, but generally we are overestimating  because by using the Dedekind zeta function we have included in the sum all the ideal classes  that might not be  principal.

\section{Some proofs of Section \ref{sec:4}}\label{sec:proof}

Lemma \ref{units} is  an  elementary corollary to \emph{Dirichlet unit theorem}, but we give a proof, as it sheds some light on the question. 
\begin{IEEEproof}[Proof of Lemma \ref{units}]
The number field  $K$ has  \emph{signature} $(0,n)$. The Dirichtlet  unit theorem then tells us that  the unit group $\OO_K^*$ has the following multiplicative structure
$$
\OO_K^*=U_{free} \times U_{roots}=\Z^{n-1}\times  U_{roots},
$$
where $U_{roots}$ is a finite torsion group containing roots of unity in $\OO_K$. Let us consider the mapping $f: \OO_K^* \to \R^{n}$
$$
u \mapsto f(u)= (\log|\sigma_1(u)|, \log|\sigma_2(u)|,\dots, \log|\sigma_n(u)|).
$$
It is well known that $f(U_{free})$ is a $(n-1)$-dimensional lattice inside $\R^{n}$.

Consider  $\psi(U_{free}) \cap B(M)$. If $\psi(u)$ happens to be inside the ball $B(M)$ of radius $M$, we have in particular that
 $|\sigma_i(u)|\leq M$ for all  $i$. It  follows that  for coordinates $\sigma_i(u)$ having absolute value greater than 1 we have $\log(|\sigma_i(u)|)\leq \log(M)$.  On the other hand if $|\sigma(u_i)|< 1$, we have that $|\log(|\sigma_i(u)|)|\leq (n-1)\log(M)$, which is a consequence of the facts that for positive coordinates $\log(|\sigma_i(u)|)\leq \log(M)$ and  $\sum_{i=1}^n \log (|\sigma_i(u)|)=0$. In summary, we have $|\log(|\sigma_i(u)|)|\leq (n-1)\log(M)$ for all $i$. 
 
Therefore,  if $\psi(u)$ is inside a ball of radius $M$, then $f(u)$ is inside a hypercube with side of length $(n-1)\log(M)$. We have that $f(U_{free})$ is a $(n-1)$-dimensional lattice, and therefore
a hypercube with  $(n-1)\log(M)$ side has less than $N \log(M)^{n-1}$ discrete elements, where $N$ is a constant independent of $M$.  It follows that
 $|\psi(U_{free}) \cap B(M)| \leq N \log(M)^{n-1}$. Now each of the elements in $\psi(U_{roots})$ is a unitary matrix. Hence for any $u= u_f u_r \in \OO_K^*$ with $u_f \in U_{free}$ and $u_r \in U_{roots}$, we have $\norm{\psi(u_f u_r)}_F=\norm{\psi(u_f)}_F$. Therefore we see that 
$$
|\psi(U_{free}\times U_{roots})\cap B(M)|=|\psi(U_{free})\cap B(M)|\cdot |U_{roots}|.
$$
It follows that
$$
|\psi(U_{free}\times U_{roots})\cap B(M)| \leq N \log(M)^{n-1}\cdot |U_{roots}|.
$$
As the group $U_{roots}$ is finite, the claim follows.
\end{IEEEproof}
 
In the following we will denote by $\mathbf{I}_K$ the set of \emph{integral ideals} of the ring $\OO_K$.

\begin{IEEEproof}[Proof of Proposition \ref{zeta}]
Using basic properties of algebraic norm and the AM-GM inequality, we have 
\[
|\det(\psi(x))|^2=|N_{K/\Q}(x)|\leq  c \cdot \norm{\psi(x)}_F^{2n},
\]
for any element $x \in X(M) \subseteq \OO_K$ and for some constant $c$. This implies 
\[
\sum_{x \in X(M)} \frac{1}{|\det (\psi(x))|^{2n_r}} \ \leq \  \sum_{\substack{x \in X\\ |N_{K/\Q}(x)|\leq c \cdot M^{2n}}} \frac{1}{|N_{K/\Q}(x)|^{n_r}},
\]
where $X$ is a  set of elements  $x\in \OO_K$, each generating a separate ideal in  $\OO_K$

From the relation between ideals and element norms we can further upper bound the above quantity by 
\[
\sum_{x \in X\atop |N_{K/\Q}(x)|\leq c \cdot M^{2n}} \frac{1}{|N_{K/\Q}(x)|^{n_r}}  \leq   \sum_{I\in \textbf{I}_K\atop |N_{K/\Q}(I)|\leq c \cdot M^{2n} }  \frac{1}{|N_{K/\Q}(I)|^{n_r}},
\]

where $\textbf{I}_K$ represents the set of all integral ideals. Note that the right-hand-side corresponds exactly to the beginning of the Dedekind zeta function at the point $n_r$. It then follows that 
\begin{multline*}
\sum_{\substack{I\in \textbf{I}_K\\ |N_{K/\Q}(I)|\leq c \cdot M^{2n} } } \frac{1}{|N_{K/\Q}(I)|^{n_r}}\\
\hspace{0.8in} \leq \left( \sum_{i<c \cdot M^{2n}, i\in \Z^+ } \frac{1}{i^{n_r}}\right)^{2n}\leq(\log(c \cdot M^{2n}))^{2n},
\end{multline*}
where the first inequality is based on a similar reasoning as in \cite[Proposition 7.2 and Corollary 3]{Nark} as well as some elementary approximation.
\end{IEEEproof}

\section{The growth of inverse determinant sums of division algebra based space-time codes}\label{sec:6}
In this section we will determine the growth of inverse determinant sums of the most well known algebraic space-time codes.
In our main results we will see that the growth of these sums, and conjecturally also their DMT, only depend on the unit groups of these algebras.

\subsection{Space-time codes from division algebras}

We now recall how to naturally build space-time lattice codes from division algebras. The algebraic results in this section are standard and can be found for example from \cite{R}.
Suppose that $F=\Q$ or $F=\Q(\sqrt{-m})$, where $m$ is a square free  natural number.
Let  $E/F$ be a cyclic field extension of degree $n$ with  Galois group $ G(E/F)= \langle\sigma\rangle $. Define a cyclic algebra
$$
\D=(E/F,\sigma,\gamma)=E\oplus uE\oplus u^2E\oplus\cdots\oplus u^{n-1}E,
$$
where   $u\in\mathcal{D}$ is an auxiliary
generating element subject to the relations
$xu=u\sigma(x)$ for all $x\in E$ and $u^n=\gamma\in (F)^*$.  We assume that $\D$ is a division algebra.

Considering $\D$ as a right  vector space over $E$, every element $x=x_0+ux_1+\cdots+u^{n-1}x_{n-1}\in\mathcal{D}$
has the following left regular representation as a matrix $\psi(x)$:
\begin{equation*}\label{esitys}
\begin{pmatrix}
x_0& \gamma\sigma(x_{n-1})& \gamma\sigma^2(x_{n-2})&\cdots &
\gamma\sigma^{n-1}(x_1)\\
x_1&\sigma(x_0)&\gamma\sigma^2(x_{n-1})& &\gamma\sigma^{n-1}(x_2)\\
x_2& \sigma(x_1)&\sigma^2(x_0)& &\gamma\sigma^{n-1}(x_3)\\
\vdots& & & & \vdots\\
x_{n-1}& \sigma(x_{n-2})&\sigma^2(x_{n-3})&\cdots&\sigma^{n-1}(x_0)\\
\end{pmatrix}.
\end{equation*}

The  mapping $\psi$ is an injective $F$-algebra homomorphism  that allows us to identify
$\D$ with its image in $M_n(\C)$.  Every non-zero element in the set $\psi(\D)\subset M_n(\C)$ is invertible, but
$\psi(\D)$ is dense and therefore not directly suitable for space-time coding.

An order of a division algebra will offer us a remedy. 
\begin{definition}
 An $\OO_F$-order $\Lambda$ in $\D$ is a subring of $\D$, having the same identity element as
$\D$, and such that $\Lambda$ is a finitely generated
module over $\OO_F$ and generates $\D$ as a linear space over
$F$.  
\end{definition}

\begin{lemma}\label{norm}
For any  element $x\in \Lambda$, we have that 
$\det(\psi(x))\in \OO_F$.
\end{lemma}

\begin{proposition}\label{Qlattice}
Suppose that $F=\Q$.
If $\Lambda$ is a  $\Z$-order in an index-$n$ division algebra $\D$, then 
$\psi(\Lambda)$ is  an $n^2$-dimensional NVD lattice in $M_n(\C)$, with
$$
\det(\psi(x))\in \Z,
$$
for all the  elements $x \in \Lambda$. 
\end{proposition}

\begin{proposition}\label{Flattice}
Suppose that $F=\Q(\sqrt{-m})$.
If $\Lambda$ is an $\OO_F$-order in an index-$n$ division algebra $\D$, then 
$\psi(\Lambda)$ is a $2n^2$-dimensional NVD lattice in $M_n(\C)$, with
$$
\det(\psi(x))\in \OO_F,
$$
for all the  elements $x \in \Lambda$. 
\end{proposition}

\begin{remark}
We note that in both cases an order $\Lambda$ is also a free $\Z$-module. This means that we have elements
$x_1,\dots,x_k\in \Lambda$ so that 
$$
\Lambda=\Z x_1 \oplus \Z x_2 \oplus \cdots \oplus \Z x_k.
$$
Therefore 
$$
\psi(\Lambda)=\Z\psi(x_1) \oplus \Z \psi(x_2) \oplus \cdots \oplus \Z\psi(x_k) \subset M_n(\C),
$$
and we can see that $\psi(x_1),\dots, \psi(x_k)$ form a basis for the lattice $\psi(\Lambda)$.
\end{remark}

The above two families cover many of the most well known codes. While we will focus only on the orders, the corresponding results hold also for principal ideals of orders (see Section \ref{divsumsecproofs}). 
For example the Perfect codes \cite{BORV} and maximal order codes \cite{VHLR}  are of the type described in  Proposition \ref{Flattice}. On the other hand the Alamouti code and the fast decodable codes in \cite{VHO} are of the type described in Proposition \ref{Qlattice}.

We now have two families of  NVD lattices with $2n^2$ and $n^2$ dimensions  in $M_n(\C)$, respectively. Below we would like to
analyze the asymptotic growth of their inverse determinant sums
$$
\sum_{ 0 \neq  x\in \Lambda \atop \norm{\psi(x)}_F\leq M} \frac{1}{|\det(\psi(x))|^{2n_r}}.
$$
The analysis will be presented in Sections \ref{complexcenter} and \ref{Qcenter} for the cases of $\Q(\sqrt{-m})$ and $\Q$, respectively. Prior to analyzing the sums, we introduce some central objects needed in both cases.

An obvious lower bound for the growth of an inverse determinant sum is given by the number of elements in the set
$$
\{ x \ : \  ||\psi(x)||_F\leq M, |\det(\psi(x))|=1, x\in \Lambda\}.
$$
This set, consisting of elements having the smallest determinant in absolute value in the lattice, can be characterized algebraically.

\begin{definition}
The unit group $\Lambda^*$ of an order $\Lambda$  consists of elements $x\in \Lambda$ such that there exists $y\in \Lambda$ with 
$xy=1_{\D}$. 
\end{definition} 
\begin{lemma}
 If the center of the algebra $\D$ is $\Q$ or a complex quadratic field, we have that
 $$
 \Lambda^*=\{ x\ : \  |\det(\psi(x))|=1, x\in \Lambda \}.
 $$
\end{lemma}

We can then write
$$
|\psi(\Lambda^*)\cap B(M)|\leq  \sum_{ 0 \neq  x\in \Lambda \atop \norm{\psi(x)}_F\leq M} \frac{1}{|\det(\psi(x))|^{2n_r}}.
$$

We still need one more object.
The following subgroup of $\Lambda^*$ will play a crucial part in the analysis of $|\psi(\Lambda^*) \cap B(M)|$.
\begin{lemma}[\cite{Kleinert} p. 211]
Suppose that the center of the algebra $\D$ is $\Q$ or  a complex quadratic field. The
 unit group $\Lambda^*$ has a subgroup
$$
\Lambda^1=\{x   \ : \  x\in \Lambda^*,  \det(\psi(x))=1\},
$$
and $[\Lambda^*:\Lambda^1]<\infty$.
\end{lemma}

The following result reveals why we are interested in the group $\Lambda^1$.
\begin{lemma}\label{norm1}
Let $\D$ be an index-$n$  $F$-central  division algebra and $\Lambda$ be an $\OO_F$ order in $\D$. We then have that
$$
 |\psi(\Lambda^1) \cap B(M)|\leq |\psi(\Lambda^*) \cap B(M)| \leq  K |\psi(\Lambda^1) \cap B(M)|, 
 $$
for some constant  $K$  that is independent of $M$.
\end{lemma}
\begin{IEEEproof}
The left side inequality is trivial. The right side    is part of the proof of Proposition \ref{Uxdiv}.
\end{IEEEproof}
In the following subsections we will see that  in the dotted sense $|\psi(\Lambda^1)\cap B(M)|$ gives not only a lower bound for the growth of the inverse determinant sum, but also  an upper bound!

\subsection{Inverse determinant sums of $\Q(\sqrt{-m})$-central division algebras}\label{complexcenter}

We first focus on the case where $\mathcal{D}$ is an index-$n$ $\Q(\sqrt{-m})$-central division algebra.

The following proposition  is an analogue to the corresponding result, Proposition \ref{numberfieldsum}, in the number field case. The proof follows similar lines.
\begin{proposition}\label{firstbound}
Let $\D$ be an index-$n$ central division algebra over $F=\Q(\sqrt{-m})$ and $\Lambda$ be an $\OO_F$-order in $\D$. Then,  for $n_r \geq n$  we have
$$
 |\psi(\Lambda^*)\cap B(M)|\leq  S_{\psi(\Lambda)}^{2n_r}(M)  \leq K (\log M)^T |\psi(\Lambda^*)\cap B(M)|,
$$
where $T$ and $K$ are constants independent of $M$.
\end{proposition}
\begin{IEEEproof}
The proof will be given in Section \ref{divsumsecproofs}.
\end{IEEEproof}

We can now see that in order to measure the asymptotic  behavior of the determinant sum it is enough to measure the growth of  $|\psi(\Lambda^*)\cap B(M)|$. However, this is not as simple a task as in the case of number fields. The unit group $\psi(\Lambda^*)$ is a wild object \cite{Kleinert}, and we need some advanced tools to solve the problem.

Lemma \ref{norm1} allows us to  consider  the asymptotic behavior of $|\psi(\Lambda^1)\cap B(M)|$  instead of the whole unit group and 
translates the problem into solvable form.

\begin{definition}\label{SLCdef}
 The set  
 $$
 \{X \,|\, X\in M_n(\C), \det(X)=1 \}
 $$
 is the Lie group $\SL_n(\C)$.
\end{definition}

The terms cocompact and discrete, appearing in the following lemma, will be explained in Section \ref{lie_sec}.
\begin{lemma}\cite[Theorem 1]{Kleinert}
Let $\D$ be an index-$n$ central division algebra over $F=\Q(\sqrt{-m})$ and $\Lambda$ an $\OO_F$ order in $\D$.
We then have that
$$
\psi(\Lambda^1)\subset {\SL}_n(\C),
$$
is a discrete \emph{cocompact} subgroup of $\SL_n(\C)$.
\end{lemma}
 The reader who is not familiar with these terms can think of an additive lattice $\Z^n$ inside of $\R^n$.  The relation between these additive groups is similar to that between the multiplicative groups $\psi(\Lambda^1)$ and  $\SL_n(\C)$. 

The previous lemma now identifies the  group $\psi(\Lambda^1)$ as a cocompact \emph{lattice} in $\SL_n(\C)$, and we can apply the 
machinery of point counting in Lie groups to prove the following. 
\begin{lemma}\label{komplexunitgroup}
Let $\D$ be an index-$n$ central division algebra over $F=\Q(\sqrt{-m})$ and $\Lambda$ an $\OO_F$-order in $\D$. We then have 
$$
|\psi(\Lambda^*)\cap B(M)| \stackrel{.}{=}|\psi(\Lambda^1)\cap B(M)|\stackrel{.}{=}  M^{2n^2-2n}.
$$
\end{lemma}
\begin{IEEEproof}
The proof  can be found in Section \ref{volume_computation}.
\end{IEEEproof}

We can now combine Proposition \ref{firstbound} and  Lemma \ref{komplexunitgroup} for the following.
\begin{thm}\label{zetadiv}
Let $\D$ be an index-$n$ $\Q(\sqrt{-m})$-central division algebra and $\Lambda$ be an $\OO_F$ order in $\D$.  Then,  for $n_r \geq n$ 
$$
S_{\psi(\Lambda)}^{2n_r}(M)\stackrel{.}{=}|\psi(\Lambda^*)\cap B(M)|\stackrel{.}{=}M^{2n^2-2n} .
$$
\end{thm}

This result reveals that asymptotically the growth of the inverse determinant sum of a division algebra-based code only depends on the unit group of the underlying order. We can also see that this growth is optimal in the sense that it meets the bound of  Proposition \ref{mainDMT}.

The analysis also reveals that all the inverse determinant sums for $\Q(\sqrt{-m})$-central division algebras have the same asymptotic behavior.  

\begin{remark}
We point out that Proposition \ref{mainDMT} already told us that the determinant sums for the algebras of this type must grow at least like $M^{2n^2-2n}$. In this section we showed that this is also an upper bound in the dotted sense.  Noting that  Proposition \ref{mainDMT} is based completely on information theory, it  is very surprising that the DMT can help to predict the distribution of norms of elements of an order. It appears that the DMT is forcing an order to have  a fairly large, that is, dense unit group.
\end{remark}

\subsection{Inverse determinant sums of $\Q$-central division algebras}\label{Qcenter}

We now concentrate on the case where the center of the division algebra is $\Q$. The most well known code of this type is the Alamouti code.
Earlier we did analyze the determinant sum of this code by observing that the sum is an Epstein zeta function, and we showed that the growth is in class
$M^0$ in  the dotted sense.  In this section we will see that this behavior is actually a particular case of a far more general theory.

Suppose that $\D$ is a $\Q$-central division algebra and  $\Lambda$ a $\Z$-order in  $\D$. Then 
$\psi(\Lambda)$ is  an  $n^2$-dimensional NVD lattice in $M_n(\C)$.

As in the previous subsection we have:
\begin{proposition}\label{firstboundZ}
Let $\D$ be an index-$n$ $\Q$-central division algebra and $\Lambda$ be a $\Z$-order in $\D$. 
Then,  for $n_r \geq n/2$  we have
$$
 |\psi(\Lambda^*)\cap B(M)|\leq  S_{\psi(\Lambda)}^{2n_r}(M)  \leq K (\log M)^T |\psi(\Lambda^*)\cap B(M)|,
$$
where $T$ and $K$ are constants independent of $M$.
\end{proposition}
\begin{IEEEproof}
The proof will be given in Section \ref{Qcenterproofs}.
\end{IEEEproof}

Similar to the previous case, we face the  problem of measuring $\abs{\psi(\Lambda^*)\cap B(M)}$, and again this reduces to measuring 
$\abs{\psi(\Lambda^1)\cap B(M)}$. Promisingly, we can again see $\Lambda^1$ as a part of $\SL_n(\C)$.
\begin{lemma}
Let $\D$ be an index-$n$ $\Q$-central division algebra and $\Lambda$ a $\Z$-order in $\D$.
We then have that
$$
\psi(\Lambda^1)\subset {\SL}_n(\C),
$$
is a discrete  subgroup of $\SL_n(\C)$.
\end{lemma}
\begin{IEEEproof}
By definition $\psi(\Lambda^1)\subset \SL_n(\C)$. It is discrete as it is a subset of a discrete set $\psi(\Lambda)$.
\end{IEEEproof}
The lacking part  here is that $\psi(\Lambda^1)$ is not "large enough" to be cocompact in  $\SL_n(\C)$ and we cannot directly employ the methods in Section \ref{lie_sec}.  Instead we have to make a detour to realize the group  $\Lambda^1$ as part of a "smaller" Lie group that will give us a tight enough fit for the ergodic methods needed for point counting in Lie groups. Unlike the case of complex quadratic centered division algebras, the structure of this algebra will have a dramatic effect on the unit group. Before proceeding, we need some definitions and results.

Consider matrices
$$
\begin{pmatrix}
A &   -B^* \\
B&  A^*                                               
\end{pmatrix}
\in M_{2n}(\C),
$$
where $*$ refers to complex conjugation and $A$ and $B$ are complex matrices in $M_n(\C)$.
We denote the set of  matrices of this type by $M_{n}(\mathbb{H})$. Indeed, there is a natural isomorphism between this ring and the ring of $n \times n$ matrices over the Hamilton quaternions $\mathbb{H}$.

\begin{definition}\label{ramified}
Suppose that $\D$ is an index-$n$  $\Q$-central division algebra. If 
$$
\D\otimes_{\Q}\R\cong M_n(\R),
$$
we say that $\D$ is not ramified at the infinite place.
If $2|n$ and 
$$
\D\otimes_{\Q}\R\cong M_{n/2}(\mathbb{H}),
$$
we say that  $\D$ is ramified at the infinite place. 
\end{definition}

\begin{lemma}\cite{R}
Suppose that $\D$ is an index-$n$  $\Q$-central division algebra. Then $\D$ has two options. Either it is ramified at the infinite place
or it is not.
\end{lemma}

If the reader is not familiar with tensoring, the main point is that there are exactly two types of $\Q$ central division algebras.
Tensoring  can then be seen as something that reveals the underlying geometric structure of the algebra.

\begin{definition}\label{SLRdef}
 The set  
 $$
 \{X \,|\, X\in M_n(\R), \det(X)=1 \}={\SL}_n(\R)
 $$
 is a   subgroup of the Lie group ${\SL}_n(\C)$. 
\end{definition}

\begin{definition}\label{SLHdef}
 The set  
 $$
 \{X \,|\, X\in M_{n/2}(\mathbb{H}), \det(X)=1 \}={\SL}_{n/2}(\mathbb{H})
 $$
 is a subgroup of the Lie group ${\SL}_n(\C)$.
\end{definition}

\begin{lemma}\label{conjugate}
Suppose we have an index-$n$ $\Q$-central division algebra $\D$ and that $\Lambda$ is a $\Z$-order in $\D$.
If $\D$ is ramified at the infinite place, there exists an invertible matrix $X\in M_n(\C)$ such that
$$
 X\psi(\Lambda^1)X^{-1}\subset {\SL}_{n/2}(\mathbb{H}).
$$

If $\D$ is not ramified at the infinite place there exists an invertible matrix $X \in M_n(\C)$, such that
$$
 X\psi(\Lambda^1)X^{-1}\subset {\SL}_n(\R).
$$
\end{lemma}
\begin{IEEEproof}
The proof will be given in Section \ref{Qcenterproofs}.
\end{IEEEproof}

In the following lemma $G$ is either ${\SL}_{n/2}(\mathbb{H})$ or ${\SL}_n(\R)$.
\begin{lemma}
Suppose that $\D$ is an index-$n$ $\Q$-central division algebra with an order $\Lambda$. If $X \in M_n(\C)$ is the matrix of Lemma \ref{conjugate}, then 
$X\psi(\Lambda_1)X^{-1}$
is a \emph{cocompact} subgroup in $G$ and
$$
 |\psi(\Lambda^1)\cap B(M)|\stackrel{.}{=} |X\psi(\Lambda^1)X^{-1}\cap B(M)|.
$$

\end{lemma}
\begin{IEEEproof}
The proof  will be given in Section \ref{Qcenterproofs}.
\end{IEEEproof}

The following lemma now follows as we can apply point counting in Lie groups to the group   $X\psi(\Lambda^1)X^{-1}$. 
Depending of the ramification at the infinite place, the counting will be done in ${\SL}_n(\R)$ or   in ${\SL}_{n/2}(\mathbb{H})$.
\begin{lemma}\label{Qcentralunitgroup}
Let $\D$ be an index-$n$ $\Q$-central division algebra and $\Lambda$ be a $\Z$-order in $\D$.
If $\D$ is ramified at the infinite place we have that
$$
|\psi(\Lambda^*)\cap B(M)| \stackrel{.}{=} M^{n^2-2n}. 
$$
If $\D$ is not ramified at the infinite place we have
$$
|\psi(\Lambda^*)\cap B(M)| \stackrel{.}{=} M^{n^2-n}. 
$$
\end{lemma}
\begin{IEEEproof}
 The proofs  will be given in  Section  \ref{volume_computation}.
\end{IEEEproof}

We can now conclude the following. 

\begin{thm}\label{Rbound}
Let $\D$ be an index-$n$ $\Q$-central division algebra  where the infinite place is not ramified and $\Lambda$  a $\Z$-order in $\D$.  Then,  for $n_r \geq n/2$  we have
$$
S_{\psi(\Lambda)}^{2n_r}(M)\dot = M^{n^2-n}.
$$
\end{thm}

\begin{thm}\label{Hbound}
Let $\D$ be an index-$n$, $2|n$,  $\Q$-central division algebra where the infinite place is ramified. Let  $\Lambda$ be a  $\Z$-order in $\D$.  Then,  for $n_r\geq n/2$ 
$$
  S_{\psi(\Lambda)}^{2n_r}(M) \dot = M^{n^2-2n}.
$$
\end{thm}

\begin{remark}
If we like to use these results in code design or in the analysis of known codes, it is crucial to recognize whether the underlying
algebra is ramified at the infinite place. We can say that  it is relatively easy. We refer the reader to \cite{VHO} for some simple methods, which will be used in the analysis of  the codes in the next example.

\end{remark}

\begin{exam}\label{Hexam}
Let us now return to the two example codes  mentioned in the introduction: $\D_2=(\Q(i)/\Q,\sigma,-3)$ and $\D_1=(\Q(i)/\Q,\sigma,3)$. 
 Both of the division algebras have  natural orders $\Lambda_i=\Z[i] \otimes u_i\Z[i]$. A straight calculation reveals that  these orders have the same geometric structure and  normalized minimum determinant.   However, these codes have drastically different inverse determinant sums. Corollary \ref{Hbound} gives growth $M^0$ for  for the code $\psi(\Lambda_2)$ and Corollary \ref{Rbound} gives growth
$M^{2}$ for the code $\psi(\Lambda_1)$.
\end{exam}

\section{Corollaries to the  DMT }\label{sec:7}
It is our belief that the inverse determinant sum of an order code derived from a division algebra indeed describes the DMT of the corresponding coding scheme for multiplexing gain $r\in [0,1]$. In this section we will turn our inverse-determinant-sum results into lower bounds for the DMT.
We will see that in the cases where the DMT of the code is known, the prediction gotten from the inverse determinant sum does give the correct result.

\begin{corollary}\label{DMTalam1}
Let $\D$ be a $\Q$-central division algebra with index $n$  and $\Lambda$ be a $\Z$-order in $\D$. Then $\psi(\Lambda)$ is an $n^2$-dimensional lattice in $M_{n}(\C)$. If $2|n$ and $\D$ is ramified at the infinite place, then the coding scheme derived from lattice $\psi(\Lambda)$  based on spherical shaping \eqref{codingscheme} achieves the DMT curve $(r,d(r))$ for $0 \leq r \leq 1$, which is a straight line connecting the points 
\begin{equation}\label{DMTlower1}
(0,nn_r) \textnormal{ and }  (1, nn_r-2n_r-n+2),
\end{equation}
when received by $n_r\geq n/2$ receiving antennas. This curve coincides with the optimal curve $d^*(r)$ for $0 \leq r \leq 1$ if and only if $n=2$ and $n_r=1$.
\end{corollary}

\begin{IEEEproof}
The DMT lower bound \eqref{DMTlower1} follows directly from Theorem  \ref{Hbound} and from an argument similar to the proof of Corollary \ref{DMTcorollary}. 
The last statement follow as the optimal DMT curve in the $n \times n_r$ channel, for $r\in [0,1]$, is a straight line connecting the points
\begin{align}
(0,nn_r) \textnormal{ and } (1,nn_r-n-n_r+1). \tag*{\IEEEQED}
\end{align}
\let\IEEEQED\relax
\end{IEEEproof}

\begin{corollary}\label{DMTreal}
Let $\D$ be a $\Q$-central division algebra with index $n$  and $\Lambda$ be a $\Z$-order in $\D$. Then $\psi(\Lambda)$ is an $n^2$-dimensional lattice in $M_{n}(\C)$. If  $\D$ is not ramified at the infinite place, then the coding scheme derived from lattice $\psi(\Lambda)$  based on spherical shaping \eqref{codingscheme} achieves the DMT curve $(r,d(r))$ for $0 \leq r \leq 1$, which is a straight line connecting the points
\begin{equation}\label{DMTlower2}
(0,nn_r) \textnormal{ and } (1, nn_r-2n_r-n+1),
\end{equation}
when received by $n_r \geq n/2$ receiving antennas. This curve  never coincides with the optimal curve.
\end{corollary}
\begin{IEEEproof}
The result follows from Theorem \ref{Rbound}.
\end{IEEEproof}

\begin{corollary}\label{DMTcomplex}
Let $\D$ be an $F$-central division algebra with index $n$ and $F=\Q(\sqrt{-m})$. If $\Lambda$ is an $\OO_F$-order inside $\D$, then $\psi(\Lambda)$ is a $2n^2$-dimensional lattice in $M_{n}(\C)$. The coding scheme derived from lattice $\psi(\Lambda)$  based on spherical shaping \eqref{codingscheme} achieves the DMT curve $(r,d(r))$ for $0 \leq r \leq 1$, which is a straight line connecting the points
$$
(0,nn_r) \textnormal{ and } (1, nn_r-n-n_r+1),
$$
when received by $n_r\geq n$ receiving antennas. It coincides with  the optimal DMT curve $d^(r)$ in the range of $0 \leq r \leq 1$ for any $n$ and $n_r$.
\end{corollary}
\begin{IEEEproof}
The result follows Theorem \ref{zetadiv}.
\end{IEEEproof}

\section{Point counting in Lie groups}\label{lie_sec}
From this section on we begin to work on proving the previously claimed results on determinant sums.  As  we saw in Section \ref{sec:6}, the growth of the inverse determinant sum of  a division algebra code 
depends essentially on the asymptotic growth of $\abs{\psi(\Lambda^1) \cap B(M)}$.  
The latter can be estimated thanks to the fact that
$\psi(\Lambda^1)$ admits a realization as a discrete cocompact subgroup of a suitable Lie group $G$. The term cocompact simply means that
the quotient group $G/\psi(\Lambda^1)$ is a compact topological space with respect to the quotient topology. 

 In this section we will present some general results on the asymptotic growth of these subgroups.   

Let  $G$ be a Lie group, where $G$ is $\SL_n(\R)$, $\SL_n(\C)$ or $\SL_n(\mathbb{H})$, and  $L$ be a discrete cocompact subgroup  of $G$.  In the following we will discuss the problem of counting the number of points of $L$
that lie inside a sphere defined with respect to the Frobenius norm. We refer the reader to \cite{GN} for the relevant definitions and an introduction to the subject. In all the statements in this section we suppose that $G$ is one of the three aforementioned Lie groups. The results are far more general, but this generality is enough for us.

Each of the groups  $G$ admits a \emph{Haar measure} that gives us a natural concept of volume $\Vol_G$.
In particular we can consider the volumes of the balls
$$
{\Vol}_G (B(M)),
$$
where $B(M)$ here refers to all the matrices in $G$ having Frobenius norm less than $M$.
 
The discrete group $L$  being cocompact in $G$ yields that the measure 
$\mu(G/L)$ induced by ${\Vol}_G$ is finite.  A general name for such group $L$ is  lattice. This is a natural generalization of an additive lattice in $\R^n$.

The following theorem is stronger than what is needed for measuring the growth of the unit group, but we need this result for the proofs of Propositions
\ref{firstbound} and  \ref{firstboundZ}.
\begin{thm}[\cite{GoNe}, Corollary 1.11 and Remark 1.12]\label{Gorodnik}
Consider a Lie group $G$, a discrete cocompact subgroup $L\subset G$ and an element $x\in G$. We then have that
$$
\lim_{M\to\infty}\left|\frac{ xL\cap B(M)}{{\Vol}_G(B(M))}\right| = K
$$
where $K$ is some nonzero constant independent of $M$. The limit  approaches $K$ uniformly for all $x\in G$.
\end{thm}
By setting $x$ to the identity matrix, one can see that the previous theorem does transform the point counting problem into an integration problem.
However, integration on a manifold  such as $\SL_n(\mathbb{H})$ is not completely straightforward.
\begin{thm}[Theorem 7.4, \cite{GW}]\label{Gorodnik2}
Suppose that $G$ is a Lie group. We then have that 
$$
{\Vol}_G(B(M))\sim M^T,
$$
for some constant $T$.
\end{thm}
The value of $T$ is well known in the  case $G= \SL_n(\R)$  and we have that $T= n^2-n$ \cite{DRS}. The corresponding results for
${\SL}_n(\mathbb{H})$ and ${\SL}_n(\C)$, although probably well-known to specialists, are not readily available in the literature, but there are general methods for calculating these asymptotic integrals (see \cite{GW} and \cite{Mau}). In order to use these methods one needs to determine some invariants of Lie algebras, related to the Lie groups under consideration. We explain these technical concepts in detail in Appendix \ref{sec:10}, where we prove that $T=2n^2-2n$ for $G=\SL_n(\C)$ and $T=4n^2-4n$ for $G=\SL_n(\mathbb{H})$, see Examples \ref{volume_SLnR}, \ref{volume_SLnH}.

\section{Proofs of section \ref{sec:6}}\label{divsumsecproofs}
In this section we suppose that the reader is familiar with algebraic number theory and the theory of central simple algebras.
One should note that we will exclusively work with \emph{maximal orders}. However, the results on the unit groups and growth of the 
inverse determinant sums hold also true for other orders. The upper bounds follow as for any order $\Lambda \subset \D$ we can find a maximal one 
$\Lambda_{max}$ such that $\Lambda \subset \Lambda_{max}$. On the other hand, the lower bound does come from the density of the unit group and the proofs work for any order.  The results to be proved work also for principal ideals of orders. In particular, our results do cover the Golden code and most of the other perfect codes. This is due to the fact that these codes have the form $A\psi(\Lambda)$, where $\Lambda$ is an order and $A$ is a matrix in $M_n(\C)$. The claim then follows from  Lemma \ref{Alattice}.

\subsection{Some preliminary algebraic results}
 Let $\D$ be an index-$n$ $F$-central division algebra and $\Lambda$  a $\Z$-order in $\D$.
The  (right) \emph{Hey zeta function} \cite{Hey} of the order $\Lambda$ is
$$
\zeta_{\Lambda}(s)=\sum_{I\in {\bf I}_{\Lambda}}\frac{1}{[\Lambda:I]^{s}},
$$
where $\Re(s)>1$ and ${\bf I}_{\Lambda}$ is the set of right ideals of $\Lambda$. When $\Re(s) >1$, this series is converging \cite{BR1}. However, we  can also consider the truncated form of this sum at the point $s=1$. We  have the following lemma.

\begin{lemma}\label{zetasum}
Let $\D$ be an index-$k$ $F$-central division algebra  and $\Lambda\subseteq \D$ be a maximal $\OO_F$-order in $\D$. If $s\geq 1$, we have that
$$
\zeta_{\Lambda}(s)(M):=\sum_{I\in {\bf I}_{\Lambda}, [\Lambda: I] \leq M}\frac{1}{[\Lambda:I]^{s}}\leq N\log(M)^K,
$$
for some constants $N$ and $K$ that are independent of $M$.
\end{lemma}
\begin{IEEEproof}
When $s>1$ the sum converges and the  bound is trivial. Let us now consider the case when $s=1$.
It first follows from \cite[p.175]{BR} where the authors 
state Hey's result 
$$
\zeta_{\Lambda}(s)=\prod_{i=0}^{k-1} \zeta_{F}(ks-i)\cdot f(s),
$$
where $\zeta_{F}$ is the Dedekind zeta function defined in \eqref{Dedekind_zeta_function}, and $f(s)$ is a function having finite Dirichlet series. In our asymptotic upper bound we  can ignore the term $f(s)$. The terms $\zeta_{F}(ks-i)$, for $i\neq k-1$, do  stay limited, when $s$ approaches $1$, and the relevant term is then
$$
\zeta_{F}(ks-k+1)\cdot g(s),
$$
where $g(s)$ has positive termed  Dirichlet series and converges at $1$.  Generally for  truncated  positive termed Dirichlet series $l_1(s)(M)$ and $l_2(s)(M)$ and for a positive real number $s$, we have that
$$
(l_1\times l_2)(s)(M)\leq l_1 (s)(M)\cdot l_2 (s)(M),
$$
where $l_1\times l_2$ is the formal product of Dirichlet series. This inequality holds even when the series do not converge.

We now have that
$$
\zeta_{\Lambda}(s)(M)\leq K\zeta_{F}(ks-k+1)(M), 
$$
where we consider $\zeta_{F}(ks-k+1)$ as a Dirichlet series, with $s$ as a variable,  and where $K$ is a constant independent of $M$.
We can now write
$$
\zeta_{F}(ks-k+1)= \sum_{n=1}^{\infty} \frac{a_n n^{k-1}}{(n^k)^s},
$$
where $a_n$ are the coefficients of  the  original Dirichlet series $\zeta_{F}(s)$.
Truncating we have
$$
\zeta_{F}(ks-k+1)(M)=\sum_{n=1}^{M^{1/k}} \frac{a_n n^{k-1}}{(n^k)^s}
$$
and in particular
$$
\lim_{s\to 1} \zeta_{F}(ks-k+1)(M)=\sum_{n=1}^{M^{1/k}}\frac{a_n}{n}.
$$
The final result now follows from Lemma \ref{zeta}.
\end{IEEEproof}

\subsection{Proofs of Subsection \ref{complexcenter}}

Let us now concentrate on the case where we have a division algebra $\D$ with a complex quadratic center $F$.
The following lemma will remind the reader of some of the previously mentioned results and state a crucial relation between the norm and index of elements in $\D$. The result  is analogous to the corresponding one in the number field case.

\begin{lemma}\cite{R}
If $\Lambda$ is a maximal $\OO_F$-order in  an index-$n$ $F$-central division algebra $\D$,
then $\psi(\Lambda)$ is a $2n^2$-dimensional $NVD$ lattice in $M_n(\C)$ and
\begin{equation}\label{normindex}
|\det(\psi(x))|^{2n}=[\Lambda:x\Lambda],
\end{equation}
where $x$ is a non-zero element of $\Lambda$.
\end{lemma}

\begin{lemma}\label{Uxdiv}
Let $\D$ be an index-$n$ $\Q(i)$-central division algebra and $\Lambda$  a $\OO_F$-order in $\D$. For any $x\neq 0 \in \Lambda$, we have
$$
 |\psi(x\Lambda^*) \cap B(M)| \leq  K |\psi(\Lambda^1)\cap B(M)|.
 $$
 for some constant  $K$, that is independent  of $x$ and  $M$.
\end{lemma}
\begin{IEEEproof}
We know that $\Lambda^1$ has a finite index inside $\Lambda^*$. Suppose that $a_1,\dots, a_j$ are some representatives   of  the cosets of the group 
$\Lambda^1$ in $\Lambda^*$. We then have that 
$$
|\psi(x\Lambda^*)\cap B(M)|=\sum_{i=1}^j |\psi(xa_i\Lambda^1)\cap B(M)|.
$$
As $|\det(\psi(xa_i))|\geq 1$, we can multiply each $\psi(xa_i)$ by a diagonal matrix $c_i I$  such that  $|c_i|\leq 1$ and $\det(c_i\psi(xa_i))=1$. Clearly
$|\psi(xa_i\Lambda^1)\cap  B(M)|\leq |c_i\psi(xa_i\Lambda^1)\cap B(M)|$ for all $i$. According to Theorem \ref{Gorodnik} we then have that
$$
|\psi(x\Lambda^*)\cap B(M)|\leq N_1|\psi(\Lambda^1)\cap B(M)|,
$$
where $N_1$ is independent of  $x$ and $M$.
\end{IEEEproof}

Now we are ready to prove that 
$$
S_{\psi(\Lambda)}^{2n_r}(M)  \stackrel{.}{=}  |\psi(\Lambda^*)\cap B(M)|.
$$

\begin{IEEEproof}[Proof of Proposition \ref{firstbound}]
From the ideal theory of orders we have that if $x\Lambda =y\Lambda$, then $x$ and $y$ must differ by a unit. Therefore we can write
$$ 
\sum_{0 \neq x\in \Lambda \atop \norm{\psi(x)}_F\leq M}\frac{1}{|\det(\psi(x))|^{2n_r}}=\sum_{x\in X(M)}\frac{|\psi(x\Lambda^*) \cap B(M)|}{|\det(\psi(x))|^{2n_r}},
$$
where $X(M)$ is a collection of non-zero elements $x \in \Lambda$, $\norm{\psi(x)}_F\leq M$, each generating a separate (right) ideal (and we suppose that $X(M)$ does include all elements in $B(M)\cap \psi(\Lambda)$) generating different ideals. According to Lemma \ref{Uxdiv}   we can upper bound the previous with
\begin{equation}\label{sum}
K\cdot|\psi(\Lambda^1) \cap B(M)|\left( \sum_{x\in X(M)}\frac{1}{|\det(\psi(x))|^{2n_r}}\right ),
\end{equation}
where $K$ is some constant independent of $M$.

Using the inequality between  Frobenius norm and determinant, we have
\[
|\det(\psi(x))|^2\leq \left(\norm{\psi(x)}_F^{2}/n \right)^n,
\]
for any element $x \in \Lambda$. Together with \eqref{normindex}, this implies that
\[
\sum_{x\in X(M)}\frac{1}{|\det(\psi(x))|^{2n_r}}=\sum_{ x \in X(M), [\Lambda:x\Lambda ]<M^{2n^2} }\frac{1}{[\Lambda:x\Lambda ]^{n_r/n}}.
\]
According to Lemma \ref{zetasum} we then have that
$$
\sum_{ x \in X(M), [\Lambda:x\Lambda ]<M^{2n^2} }\frac{1}{[\Lambda:x\Lambda ]^{n_r/n}}\leq K(\log(M))^T,
$$ 
 where $T$ and $K$ are some  constants independent of $M$.
The final result now follows by substituting this  into \eqref{sum}.
\end{IEEEproof}

\subsection{Proofs of Subsection \ref{Qcenter}}\label{Qcenterproofs}
The reader shall notice that in order to keep  Section \ref{Qcenter} as simple as possible we did not reveal  how  we actually prove the 
given results. However, the proofs of all the results given in Section \ref{Qcenter} can be easily derived from the results of this section.

Suppose that $\D$ is an index-$n$  $\Q$-central division algebra and  $\Lambda\subset \D$ a $\Z$-order.  We are now interested in the behavior of the determinant sum
$$
\sum_{ 0 \neq x\in \Lambda \atop \norm{\psi(x)}_F\leq M} \frac{1}{|\det(\psi(x))|^{2n_r}}.
$$

However, unlike in the case where the center is complex quadratic, we cannot approach the problem directly. Instead we will use another, geometrically more revealing, embedding $\psi_{abs}:\D  \mapsto M_n(\C)$ (to be defined later) and we will study the corresponding sum
$$
\sum_{0 \neq x\in \Lambda \atop \norm{\psi_{abs}(x)}_F\leq M} \frac{1}{|\det(\psi_{abs}(x))|^{2n_r}}.
$$
In the end we will prove that  the behavior of this sum completely describes the behavior of the original sum, too.

Recall that there are exactly two options for a $\Q$-central division algebra $\D$, either
$$
\D\otimes_{\Q}\R\cong M_n(\R) \subset M_n(\C)
$$
or
$$
\D\otimes_{\Q}\R\cong M_{n/2}(\mathbb{H}) \subset M_n(\C).
$$
In both cases we will denote the  corresponding  isomorphisms by $\psi_{abs}$. With abuse of notation we define $\psi_{abs}(x)=\psi_{abs}(x \otimes 1)$ for $x \in \D$.

\begin{lemma}\cite{R}
If $\Lambda$ is a maximal $\Z$-order in  an index-$n$ $\Q$-central division algebra $\D$,
then $\psi_{abs}(\Lambda)$ is an $n^2$-dimensional NVD lattice in $M_n(\C)$ and
\begin{equation}\label{normindexQ}
|\det(\psi_{abs}(x))|^{n}=[\Lambda:x\Lambda].
\end{equation}
\end{lemma}

Now we can proceed just as in the case of division algebra with complex quadratic center.

\begin{proposition}\label{absZ}
Let $\D$ be an index-$n$ $\Q$-central division algebra and $\Lambda$ be a  maximal $\Z$-order in $\D$.  Then,  for $n_r \geq n/2$  we have
\begin{align*}
 &|\psi_{abs}(\Lambda^1)\cap B(M)|\leq  S_{\psi_{abs}(\Lambda)}^{2n_r}(M) \\
 & \leq K (\log(M))^T |\psi_{abs}(\Lambda^1)\cap B(M)|
\end{align*}
where $T$ and $K$ are constants independent of $M$.
\end{proposition}

In the following  we will denote ${\SL}_n(\mathbb{H})$ and ${\SL}_n(\R)$ by $G$.
Just as in the case of ${\SL}_n(\C)$,  we have the following.
\begin{lemma}\cite[Theorem 1]{Kleinert}
Suppose that $\D$ is a $\Q$-central division algebra and  $\Lambda$  a $\Z$-order in $\D$. We then have that
$$\psi_{abs}(\Lambda^1)\subseteq G$$
is a cocompact lattice in $G$.
\end{lemma}
 
Using  point counting in Lie groups, we now have:
\begin{proposition}\label{zetadiv2}
Let $\D$ be an index-$n$  $\Q$-central division algebra  and  $\Lambda$  a maximal $\Z$-order in $\D$.    Then,  for $n_r\geq n/2$ 
$$
S_{\psi_{abs}(\Lambda)}^{2n_r}(M)\dot = {\Vol}_G(B(M)).
$$
\end{proposition}

According to Examples \ref{volume_SLnR} and \ref{volume_SLnH} we now have the following desired results.
\begin{corollary}\label{Rboundabs}
Let $\D$ be an index-$n$ $\Q$-central division algebra  where the infinite place is not ramified and $\Lambda$  a  maximal $\Z$-order in $\D$.  Then,  for $n_r \geq n/2$  we have
$$
S_{\psi_{abs}(\Lambda)}^{2n_r}(M)\dot = M^{n^2-n}.
$$
\end{corollary}

\begin{corollary}\label{Hboundabs}
Let $\D$ be an index-$n$, $2|n$,  $\Q$-central division algebra where the infinite place is ramified. Let  $\Lambda$ be a maximal $\Z$-order in $\D$.   Then,  for $n_r\geq n/2$ 
$$
  S_{\psi_{abs}(\Lambda)}^{2n_r}(M) \dot = M^{n^2-2n}.
$$
\end{corollary}
We are now ready to return to the original embedding $\psi$ of the division algebra.

\begin{lemma}\label{conjugate2}
Suppose that $\D$ is an index-$n$ $\Q$-central division algebra and that $\Lambda$ is a $\Z$-order in $\D$. Then there exists $A \in M_n(\C)$ such that
$$
\psi(x)=A^{-1}\psi_{abs}(x)A,
$$
for every element $x\in \D$.
\end{lemma}
\begin{IEEEproof}
We can build a well defined  mapping $$f:\D\otimes_{\Q}\C \to M_n(\C),$$ where   $f(d\otimes c )=\psi_{abs}(d) \cdot c\mathbf{I}$ and $\mathbf{I}$ is the identity matrix. It is then easy to prove that this is a bijective $\C$-algebra homomorphism.

We also have a $\C$-algebra  morphism $g: \D\otimes_{\Q}\C  \to  M_n(\C)$, where $g(d\otimes c)=\psi(d)\cdot c\mathbf{I}$. This is just as well a bijection.   The \emph{Skolem Noether theorem}   now states  that there exists an invertible matrix $A\in M_n(\C)$, such that $f(x)=Ag(x)A^{-1}$ for every element $x$ in $\D\otimes_{\Q}\C$. In particular, we have that $\psi_{abs}(d)=f(d\otimes 1_{\C})=Ag(d\otimes 1_{\C})A^{-1}=A\psi(d)A^{-1}.$
\end{IEEEproof}

\begin{proposition}\label{equivalence}
Suppose that $\D$ is a $\Q$-central division algebra and that $\Lambda$ is a $\Z$-order in $\D$ we then have that
$$
S_{\psi_{abs}(\Lambda)}^{2n_r}(M)\dot= S_{\psi(\Lambda)}^{2n_r}(M).
$$
\end{proposition}
\begin{IEEEproof}
Combining Lemma \ref{conjugate2} and   Proposition \ref{Alattice} gives us this result.
\end{IEEEproof}

Theorems \ref{Hbound} and \ref{Rbound} now directly follow from  Corollaries \ref{Rboundabs} and \ref{Hboundabs}

\section{Concluding remarks and suggestions for further work}

In this paper we laid a basis for studying inverse determinant sums and developed methods for analyzing inverse determinant sums and DMTs of large families of algebraic codes.  We introduced several techniques, not used before in algebraic space-time coding, and revealed surprisingly tight connections between information theoretic and algebraic concepts.

There are now several directions where this study can be continued. Let us  shortly describe few of them.
 The most straightforward problem  is the tightening of  the results we have gotten, so that we can make a difference between codes that in the rough asymptotic sense, we have mostly discussed,  are similar. Preliminary research suggests that our methods can be sharpened to consider  also sums $\tilde{S}_L^m(M)$, introduced in Section \ref{basicprob}.   Can these  more refined methods then be used to find the division algebras that yield the optimal growth for corresponding sums $\tilde{S}_{\psi(\Lambda)}^m(M)$?

It seem to be that  the growth of an inverse determinant sum always describes  the DMT of a minimum delay space-time code for multiplexing gains $r\in[0,1]$. Can this be proved or disproved? Can one give a more direct proof for the results in \ref{mathematicalcomment}?

\begin{appendices}

\section{Complex and real Lie algebras, root systems and highest weight} \label{sec:10}
The aim of this appendix is to compute the constant $T$ in Theorem \ref{Gorodnik2} when the group $G$ is $\SL_n(\C)$, $\SL_n(\R)$ or $\SL_{n/2}(\mathbb{H})$.\\
In order to do so, we need some basic facts about Lie algebras. For a general introduction to Lie algebras and beyond, we refer the reader to \cite{He,Kn}. \\
A (finite dimensional) \emph{Lie algebra} over the field $\F$ is a finite dimensional vector space $\mathfrak{g}$ over $\F$ endowed with a bilinear product $[\cdot,\cdot]:\mathfrak{g} \times \mathfrak{g} \to \mathfrak{g}$, called the Lie bracket, such that 
$$\forall x \in \mathfrak{g}, \quad [x,x]=0$$
and satisfying the Jacobi identity 
$$\forall x,y,z \in \mathfrak{g}, \quad [[x,y],z]+[[y,z],x]+[[z,x],y]=0.$$
For any Lie algebra $\mathfrak{g}$, we can define a mapping $\ad: \mathfrak{g} \to \End_{\F}\mathfrak{g}$ such that $\forall x, y \in \mathfrak{g}$, $(\ad x)(y)=[x,y]$, and a bilinear form (\emph{Killing form}) $k$ on $\mathfrak{g}$ given by 
$k(x,y)=\tr(\ad x\ad y)$. We will only consider the case where $\F$ is equal to $\R$ or $\C$. In this case, $\mathfrak{g}$ is \emph{semisimple} if the Killing form is non-degenerate. We will say that $\mathfrak{g}$ is \emph{abelian} if $\Ker(\ad)=\mathfrak{g}$, or equivalently,
$$\forall x,y \in \mathfrak{g}, \quad [x,y]=0.$$ 
Even though we are mainly interested in real Lie algebras, it will be easier to define the notions of root system and weights in the case of complex Lie algebras and then derive the corresponding definitions for the real case. \\
\emph{Notation:} We denote by $\{E_{ij}\}$ the standard basis of $M_n(\C)$ and by $\{e_{ij}\}$ the corresponding dual basis. To simplify notation, we write $E_i=E_{ii}$ and $e_i=e_{ii}$. We will always suppose that $n>1$ in the sequel. 

\subsection{Root space decomposition and irreducible representations of complex Lie algebras}
Let $\mathfrak{g}$ be a semisimple Lie algebra over $\C$. A \emph{Cartan subalgebra} $\mathfrak{h}$ is a maximal abelian subalgebra such that $\forall h \in \mathfrak{h}$, $\ad h$ is diagonalizable. Given a Cartan subalgebra $\mathfrak{h}$, let $\mathfrak{h}^*$ be its dual as a vector space. \\

For $\alpha \in \mathfrak{h}^*\setminus\{0\}$, let
$$\mathfrak{g}_{\alpha}=\{x \in \mathfrak{g} \;|\; [h,x]=\alpha(h)x \quad \forall h \in \mathfrak{h}\}.$$
If $\mathfrak{g}_{\alpha} \neq \{0\}$, we say that $\alpha$ is a \emph{root} of $(\mathfrak{g},\mathfrak{h})$ (or simply a root of $\mathfrak{g}$ with abuse of notation).  We denote the set of all roots of $\mathfrak{g}$ by $\Phi$. The following \emph{root space decomposition} holds:
\begin{equation} \label{root_space_decomposition}
\mathfrak{g}=\mathfrak{h} \oplus \bigoplus_{\alpha \in \Phi} \mathfrak{g}_{\alpha}.
\end{equation}
Consider the $\R$-vector space
\begin{equation}
\mathfrak{h}(\R)=\{h \in \mathfrak{h} \; |\; \alpha(h) \in \R \quad \forall \alpha \in \Phi\} \label{h_R}.
\end{equation}
One can show that $\mathfrak{h}=\mathfrak{h}(\R) \oplus i \mathfrak{h}(\R)$, so that an basis of $\mathfrak{h}(\R)$ over $\R$ is also a basis of $\mathfrak{h}$ over $\C$. Every choice of an ordered basis $\{h_1,\ldots,h_r\}$ of $\mathfrak{h}(\R)$ induces a partition of the roots into positive and negative roots as follows. \\
Given a root $\alpha \in \Phi$, we write $\alpha>0$ if $\exists k \leq r$ such that $\alpha(h_i)=0$ for $1 \leq i \leq k-1$ and $\alpha(h_k)>0$, and $\alpha<0$ otherwise \cite{Kn}. We denote the set of \emph{positive roots} by $\Phi^+$.
A positive root $\alpha \in \Phi^+$ is called \emph{simple} if it cannot be written as a sum of positive roots. We denote the set of simple roots by $\Delta$.
\smallskip \par 
Now consider a complex \emph{representation} of $\mathfrak{g}$, that is a morphism $\mathfrak{g} \to \mathfrak{gl}(V)$ where $V$ is a finite-dimensional complex vector space. Here $\mathfrak{gl}(V)=\End(V)$ viewed as a Lie algebra with the commutator $[f,g]=fg-gf$ as Lie bracket.\\
A \emph{subspace} $W \subset V$ is \emph{invariant} under the representation $\rho$ if $\forall x \in \mathfrak{g}$, $\rho(x)(W) \subseteq W$. The representation $\rho$ is \emph{irreducible} if $V$ does not contain any nontrivial invariant subspace.\\
Given $\lambda \in \mathfrak{h}^*$, we define 
$$V_{\lambda}=\{v \in V \;|\; \rho(h)v=\lambda(h)v \quad \forall h \in \mathfrak{h}\}.$$
If $V_{\lambda} \neq \{0\}$ we say that $\lambda$ is a \emph{weight}. Let $A_{\rho}$ be the set of weights: then we have the weight space decomposition
$$V=\bigoplus_{\lambda \in A_\rho}V_{\lambda}.$$ 
A \emph{highest weight vector} is a nonzero vector $v_{\lambda}$ that belongs to some weight space $V_{\lambda}$ and such that $\forall \alpha \in \Phi^+$, $\forall x_{\alpha} \in \mathfrak{g}_{\alpha}$, $\rho(x_{\alpha})v_{\lambda}=0$.
In this case $\lambda$ is called a \emph{highest weight}.\\
It can be shown that every finite-dimensional representation of a semi-simple Lie algebra $\mathfrak{g}$ admits a highest weight vector; the highest weight vectors of an irreducible representation of $\mathfrak{g}$ are unique up to multiplication by nonzero scalars. Equivalently, the highest weight is unique and the corresponding weight space is one-dimensional. 

\begin{exam}[\emph{$\mathfrak{sl}_n(\C)$ as a complex Lie algebra}] \label{SLnC_complex}
The complex Lie algebra corresponding to the Lie group $G=\SL_n(\C)$ is 
$$\mathfrak{sl}_n(\C)=\{X \in M_n(\C) \; | \; \tr(X)=0\}$$
with the Lie bracket $[X,Y]=XY-YX$. One can show that it is semisimple; the set of trace zero diagonal matrices $\mathfrak{h}$ is a Cartan subalgebra; it is a vector space of dimension $n-1$ over $\C$. We choose the ordered basis $\{E_1-E_n,\ldots,E_{n-1}-E_n\}$ of $\mathfrak{h}$.\\ 
Note that for $\mathfrak{sl}_n(\C)$, if we consider two elements $H=a_1E_1+\ldots+a_nE_n$, $H'=a_1'E_1+\ldots+a_n'E_n \in \mathfrak{h}$, we have $\forall i,j \in \{1,\ldots,n\}$,
\begin{equation} \label{bracket}
[H,E_{ij}]=(a_i-a_j)E_{ij},
\end{equation}
so $\ad(H)$ is diagonal with diagonal elements $a_i-a_j$, and 
$$k(H,H')=\sum_{i\neq j}(a_i-a_j)(a_i'-a_j')=2n\tr(HH').$$  
It is not hard to see that the set of roots is 
$$\Phi=\{e_i-e_j \,|\, i \neq j\}.$$
In fact, from (\ref{bracket}) we find that $\forall i \neq j$, $\C E_{ij}$ is contained in the root space $\mathfrak{g}_{\alpha}$ with $\alpha=e_i-e_j$. By (\ref{root_space_decomposition}), all the root spaces $\mathfrak{g}_{\alpha}$ are one-dimensional and of the form $\C E_{ij}$. 
Moreover,
\begin{itemize}
\item[-] the set of positive roots is 
\begin{equation}
\Phi^+=\{e_i-e_j\;|\; 1 \leq i<j \leq n\} \label{positive_roots}
\end{equation}
\item[-] the set of simple roots is 
$$\Delta=\{e_i-e_{i+1}\;|\; 1 \leq i \leq n-1\}.$$
\end{itemize}
Now consider the irreducible representation $\rho$ over $V=\C^n$ induced by the usual multiplication of matrices by vectors.\\
If $v=(v_1,\ldots,v_n) \in \C^n \setminus \{0\}$ is a highest weight vector for this representation, then for $1\leq i<j \leq n$, $\forall \alpha=e_i-e_j \in \Phi^+$, and $\forall x_{\alpha} \in \mathfrak{g}_{\alpha}=\C E_{ij}$, we must have $\rho(x_{\alpha})v=E_{ij}v=0$. Consequently, $(E_{ij}v)_i=v_j=0$ $\forall j>1$. So the only possible highest weight vector (up to multiplication by scalars) is $v_{\lambda}=(1,0,\ldots,0)$. The corresponding $\lambda$ must satisfy 
$0=(E_i-E_n)v=\lambda(E_i-E_n)v$ \; $\forall 1<i<n$ and $v=(E_1-E_n)v=\lambda(E_1-E_n)v$. Therefore the highest weight is $\lambda_1=e_1-e_n=(e_1)_{|\mathfrak{h}^*}$.
\end{exam}

\subsection{Real Lie algebras, restricted root systems and restricted weights} \label{real_Lie_algebras}
Up to now we have only considered complex Lie algebras; however, a complex Lie algebra $\mathfrak{g}$ can also be viewed as a real Lie algebra by restriction of scalars. In this case we will denote it by $\mathfrak{g}_{\R}$, the \emph{realification} of $\mathfrak{g}$. 
On the other hand, given a real Lie algebra $\bar{\mathfrak{g}}$ we can define its \emph{complexification} $\bar{\mathfrak{g}}(\C)=\bar{\mathfrak{g}} \oplus i \bar{\mathfrak{g}}$, which is again a real Lie algebra with the following extension of the Lie bracket: $\forall x_1,x_2,y_1,y_2 \in \mathfrak{g}$,
$$ [x_1+ix_2,y_1+iy_2]=([x_1,y_1]-[x_2,y_2])+i([x_1,y_2]+[x_2,y_1]).$$ 
We will say that the real Lie subalgebra $\bar{\mathfrak{g}}$ of the complex Lie algebra $\mathfrak{g}$ is a \emph{real form} of $\mathfrak{g}$ if $\mathfrak{g}_{\R}=\bar{\mathfrak{g}}(\C)$ \cite{On}.
\medskip \par
We will thus consider real Lie algebras $\bar{\mathfrak{g}}$ that fall into these two main cases:
\begin{enumerate}
\item[a)] $\bar{\mathfrak{g}}=\mathfrak{g}_{\R}$ is the realification of a complex Lie algebra $\mathfrak{g}$;
\item[b)] $\bar{\mathfrak{g}}$ is a real form of a complex Lie algebra $\mathfrak{g}$.
\end{enumerate}
We start by focusing on the second case. The main reference for this section is \cite{On}.

\subsubsection*{Real forms of complex Lie algebras}
Real forms are better understood by studying the corresponding real structures.\\ 
A \emph{real structure} of a complex Lie algebra $\mathfrak{g}$ is an anti-involution $\sigma: \mathfrak{g} \to \mathfrak{g}$, that is, an $\R$-linear map $\sigma: \mathfrak{g} \to \mathfrak{g}$ such that $\forall x \in \mathfrak{g}, \;\sigma(ix)=-i\sigma(x)$.
Given a real structure $\sigma$ of $\mathfrak{g}$, the $\R$-subalgebra $\mathfrak{g}^{\sigma}$ of its fixed points is a real form of $\mathfrak{g}$; conversely, every real form is the fixed subalgebra of some real structure.\\
A complex semisimple Lie algebra always admits a real structure $\tau$ such that the restriction of the Killing form to $\mathfrak{g}^{\tau}$ is negative definite; in this case, $\mathfrak{u}=\mathfrak{g}^{\tau}$ is called a \emph{compact real form}.\\
Any involutive automorphism $\theta$ of $\mathfrak{g}$ commuting with $\tau$ determines a real structure $\sigma=\tau\theta$. Since $\theta^2$ is the identity, $\theta$ is diagonalizable with eigenvalues $\pm1$, and considering the corresponding eigenspaces we obtain a decomposition $\mathfrak{g}=\mathfrak{g}_{+} \oplus \mathfrak{g}_{-}$, where $\mathfrak{g}_{\pm}=\{x \in \mathfrak{g} \,|\, \theta(x)=\pm x\}$. If $\bar{\mathfrak{g}}=\mathfrak{g}^{\sigma}$, let $\bar{\mathfrak{g}}_{\pm}=\mathfrak{g}_{\pm} \cap \bar{\mathfrak{g}}$. \\
We say that a decomposition $\bar{\mathfrak{g}}=\mathfrak{k} \oplus \mathfrak{p}$ is a \emph{Cartan decomposition} if 
$[\mathfrak{k},\mathfrak{k}] \subseteq \mathfrak{k},\quad [\mathfrak{k},\mathfrak{p}]\subseteq \mathfrak{p},\quad [\mathfrak{p},\mathfrak{p}]\subseteq \mathfrak{k}$
and
$k(x,x)<0 \quad \forall x \in \mathfrak{k} \setminus \{0\}, \quad k(x,x)>0 \quad \forall x \in \mathfrak{p} \setminus \{0\}$.\\
It is not hard to see that the decomposition given by $\mathfrak{k}=\bar{\mathfrak{g}}_{+}$, $\mathfrak{p}=\bar{\mathfrak{g}}_{-}$
is indeed a Cartan decomposition of $\bar{\mathfrak{g}}$. \\
Consider a maximal commutative subalgebra $\mathfrak{a}$ in $\mathfrak{p}$. One can show that all such subalgebras have the same dimension $l$, called the \emph{real rank} of $\bar{\mathfrak{g}}$ \cite{VK}. \\
Let $\bar{\mathfrak{h}}$ be a maximal commutative subalgebra of $\bar{\mathfrak{g}}$ containing $\mathfrak{a}$. Then one can show that
 $\mathfrak{a}=\bar{\mathfrak{h}} \cap \mathfrak{p}$ \cite{He}. Clearly we have the decomposition $\bar{\mathfrak{h}}=\mathfrak{a} \oplus \mathfrak{t}$, where $\mathfrak{t}=\bar{\mathfrak{h}} \cap \mathfrak{k}$. 
Moreover, the complexification $\mathfrak{h}=\bar{\mathfrak{h}} \oplus i\bar{\mathfrak{h}}$ is a Cartan subalgebra of $\mathfrak{g}$ \cite{He}.
In this case, we say that $\bar{\mathfrak{h}}$ is a Cartan subalgebra of $\bar{\mathfrak{g}}$. \\
It is not hard to see that $\mathfrak{h}(\R)$ defined in (\ref{h_R}) is another real form of $\mathfrak{h}$; one can show that $\mathfrak{h}(\R)=\mathfrak{a}\oplus i\mathfrak{t}$. Since every root $\alpha$ in the set $\Phi$ of roots of $\mathfrak{g}$ is real-valued on $\mathfrak{h}(\R)$, by choosing a suitable ordered basis of $\mathfrak{h}(\R)$ we can obtain a new partition of $\Phi$ into positive and negative roots. It is essential to choose a \vv{smart} ordered basis which is \vv{compatible} with $\mathfrak{a}$, in the sense that a root such that its restriction to $\mathfrak{a}$ is positive must also be positive. For example, we can choose an ordered basis of $\mathfrak{a}$ followed by an ordered basis of $i\mathfrak{t}$ \cite{He}.\\
Consider the set of positive roots $\Phi^+$ of $\mathfrak{g}$ with respect to this basis. Let $\Phi_c^+$ be the subset of positive roots which vanish on $\mathfrak{a}$ (also called \emph{compact roots}) and let $\Phi_{nc}^{+} =\Phi^{+}\setminus \Phi_{c}^{+}$ (\emph{non-compact roots}). Then we can obtain the set $\bar{\Phi}^+$ of \emph{positive restricted roots} of $(\bar{\mathfrak{g}},\mathfrak{a})$ with their multiplicities by restricting the roots in $\Phi_{nc}^{+}$ to $\mathfrak{a}$. \\
The \emph{simple restricted roots} $\bar{\Delta}$ are defined from the positive restricted roots in a similar way to the complex Lie algebra case. \\
From the root space decomposition of $\mathfrak{g}$ we can thus obtain a restricted root space decomposition  
$$\bar{\mathfrak{g}}=\bar{\mathfrak{h}} \oplus \sum_{\gamma \in \bar{\Phi}} \bar{\mathfrak{g}}_{\gamma},$$
where the restricted root spaces $\bar{\mathfrak{g}}_{\gamma}$ are given by
$$ \bar{\mathfrak{g}}_{\gamma}=\bar{\mathfrak{g}} \cap \left( \sum_{\alpha \in \Phi, \; \alpha_{|\mathfrak{a}}=\gamma} \mathfrak{g}_{\alpha}\right).$$
The \emph{multiplicity} of the restricted root $\gamma$ is $m_{\gamma}=\dim_{\R} \bar{\mathfrak{g}}_{\gamma}$.
\subsubsection*{Irreducible real representations of real Lie algebras}
Let $\bar{\rho}$ be a representation of a real Lie algebra $\bar{\mathfrak{g}}$ over a real vector space $\bar{V}$. Then 
we can extend $\bar{\rho}$ to a complex representation $\bar{\rho}^{\C}: \bar{\mathfrak{g}} \to \mathfrak{gl}(V)$ on the complexification $V$ of $\bar{V}$. 
If $\bar{\rho}^{\C}$ is irreducible, then $\bar{\rho}$ is also irreducible, but the opposite is not necessarily true.\\ 
Conversely, given a complex representation $\rho: \bar{\mathfrak{g}} \to \mathfrak{gl}(V)$ of the real Lie algebra $\bar{\mathfrak{g}}$ on the complex vector space $V$, we can regard $\rho$ as a real representation $\rho_{\R}$ of $\bar{\mathfrak{g}}$ on the realification $V_{\R}$ of $V$. If ${\rho}_{\R}$ is irreducible, then $\rho$ also is. \\
Furthermore, it can be proven \cite{On} that given $\bar{\rho}: \bar{\mathfrak{g}} \to \mathfrak{gl}(\bar{V})$ irreducible, then either
\begin{enumerate}
\item[i)] $\bar{\rho}^{\C}$ irreducible, or
\item[ii)] $\bar{\rho}=\rho_{\R}$ where $\rho$ is an irreducible complex representation. 
\end{enumerate} 
In either case, we denote by $\rho(\C)$ the corresponding complex representation of $\mathfrak{g}$ (the straightforward extension to $\mathfrak{g}$ of ${\bar{\rho}}^{\C}$ in case (i), and of $\rho$ in case (ii)), which turns out to be irreducible too.  \\
The \emph{restricted weight spaces} of $\bar{\rho}$ are defined in a similar way to the weight spaces \cite{GJT}: for $\lambda \in \mathfrak{a}^*$, we can set
$$\bar{V}_{\lambda}=\{v \in \bar{V} \; |\; \bar{\rho}(x)v=\lambda(x) \;\; \forall x \in \mathfrak{a}\}$$
The restricted weight subspace $\bar{V}_{\lambda}$ of $\bar{\rho}$ is the sum of the weight subspaces of $\rho(\C)$ corresponding to $\mu \in \mathfrak{h}^*$ such that $\mu_{|\mathfrak{a}}=\lambda$. The restriction to $\mathfrak{a}$ of the highest weight of $\rho(\C)$ is the highest restricted weight of $\bar{\rho}$.

\begin{exam}[\emph{$\mathfrak{sl}_n(\R)$ as a real form of $\mathfrak{sl}_n(\C)$}] \label{SLnR}
The involution $\tau(X)=-X^H$ of the complex Lie algebra $\mathfrak{g}=\mathfrak{sl}_n(\C)$ gives rise to a compact real form. \\
The involutive automorphism $\theta(X)=-X^t$ which commutes with $\tau$ determines the real structure $\sigma(X)=\tau \theta (X)=X^*$, which corresponds to real form $\bar{\mathfrak{g}}=\mathfrak{g}^{\sigma}=\mathfrak{sl}_n(\R)$ of real matrices with trace zero.
The involution $\theta$ can be used to define a Cartan decomposition into symmetric matrices $\mathfrak{p}=\bar{\mathfrak{g}}_{-}=\left\{X \in \mathfrak{sl}_n(\R)\;|\; X^t=X\right\}$ and antisymmetric matrices $\mathfrak{k}=\bar{\mathfrak{g}}_{+}=\left\{X \in \mathfrak{sl}_n(\R)\;|\; X^t=-X\right\}$.\\
The set $\mathfrak{a}$ of real diagonal matrices with trace zero is clearly a commutative subalgebra of $\mathfrak{p}$ of dimension $n-1$ and it is not hard to see that it is maximal, so that referring to the notation of Section \ref{real_Lie_algebras} we can choose $\bar{\mathfrak{h}}=\mathfrak{h}(\R)=\mathfrak{a}$ and $\mathfrak{t}=\{0\}$. The set of positive roots $\Phi^+$ of $\mathfrak{sl}_n(\C)$ defined in (\ref{positive_roots}) is compatible with $\mathfrak{a}$. None of the positive roots vanish on $\mathfrak{a}$, therefore 
\begin{align*}
&\bar{\Phi}^+=(\Phi_{nc}^+)_{|\mathfrak{a}}=\Phi^+_{|\mathfrak{a}}=\{(e_i-e_j)_{|\mathfrak{a}}\;|\; i<j\},\\
&\bar{\Delta}=\{(e_i-e_{i+1})_{|\mathfrak{a}}\;|\; 1 \leq i \leq n-1\}
\end{align*}
(see Example \ref{SLnC_complex}). The real representation on $\R^n$ induced by matrix-vector multiplication is irreducible and the highest weight is $(e_1)_{|\mathfrak{a}}$ (this corresponds to case (i) in the previous discussion about real irreducible representations). 
\end{exam}

\begin{exam}[\emph{$\mathfrak{sl}_{\frac{n}{2}}(\Hh)$ as a real form of $\mathfrak{sl}_n(\C)$}] \label{SLnH}
If $n=2m$ is even, we consider the real structure $\sigma$ of $\mathfrak{sl}_{2m}(\C)$ given by $\sigma(X)=-SX^*S$, where
$$S=\begin{pmatrix} 0 & -I_m \\ I_m & 0\end{pmatrix}.$$
The set of fixed points of $\sigma$ is the real Lie algebra 
$$\mathfrak{sl}_m(\Hh)=\left\{ \begin{pmatrix} A & -B^* \\ B & A^*\end{pmatrix} \; |\; A,B \in M_m(\C), \; \Re(\tr(A))=0\right\},$$    
which is a real form of $\mathfrak{sl}_{2m}(\C)$, since 
$\mathfrak{sl}_{2m}(\C)=\mathfrak{sl}_m(\Hh) \oplus \mathfrak{sl}_m(\Hh)i$. 
Its dimension as a real vector space is clearly $4m^2-1$. 
Referring to the terminology defined before, and considering the compact real structure $\tau(X)=-X^H$ of $\mathfrak{sl}_{2m}(\C)$, and the anti-involution $\theta(X)=\sigma\tau(X)=SX^tS$, we get the Cartan decomposition $\mathfrak{sl}_m(\Hh)=\mathfrak{k} \oplus \mathfrak{p}$ with respect to $\theta$, with
\begin{align*}
&\mathfrak{k}=\left\{ \begin{pmatrix} A & -B^* \\ B & A^*\end{pmatrix} \; |\; A^H=-A, B^t=B\right\},\\
&\mathfrak{p}=\left\{ \begin{pmatrix} A & -B^* \\ B & A^*\end{pmatrix} \; |\; A^H=A, B^t=-B,  \Re(\tr(A))=0\right\}
\end{align*}
It is not hard to see that
$$\mathfrak{a}=\{\diag(a_1,\ldots,a_m,a_1,\ldots,a_m) \; | \; a_i \in \R,\; \sum_{i=1}^m a_i=0\}$$
is a maximal commutative subalgebra of $\mathfrak{p}$ of dimension $m-1$ over $\R$, and 
\begin{align*}
&\bar{\mathfrak{h}}=\left\{ \diag(h_{1},\ldots,h_{m},h_{1}^*,\ldots,h_{m}^*) \; |\; \Re\left(\sum_{i=1}^m h_{i}\right)=0\right\}
\end{align*}
is a maximal commutative subalgebra of $\bar{\mathfrak{g}}=\mathfrak{sl}_m(\Hh)$ containing $\mathfrak{a}$, of dimension $2m-1$ over $\R$. Moreover, keeping the notation of Section \ref{real_Lie_algebras}, we have
$$\mathfrak{t}=\bar{\mathfrak{h}} \cap \mathfrak{k}=\{\diag(ib_1,\ldots,ib_m,-ib_1,\ldots,-ib_m) \; | \; b_i \in \R\},$$
which has dimension $m$ over $\R$, and $\mathfrak{h}(\R)=\mathfrak{a}\oplus i\mathfrak{t}$ is the set of diagonal matrices in $\mathfrak{sl}_{2m}(\R)$.\\
Let's now consider an ordered basis of $\mathfrak{h}(\R)$ which is compatible with $\mathfrak{a}$, for example the one consisting of the basis $\{E_i-E_m+E_{i+m}-E_{2m}\}_{1 \leq i \leq m-1}$ of $\mathfrak{a}$ followed by the basis $\{E_i-E_{i+m}\}_{1 \leq i \leq m}$ of $i\mathfrak{t}$. 
Recall that the $2m(2m-1)$ roots of $\mathfrak{g}=\mathfrak{sl}_{2m}(\C)$ are given by 
$$\Phi=\{e_i-e_j, \; i \neq j, \; 1 \leq i, j \leq 2m\}.$$
With the chosen ordering, it is not hard to see that the set of positive roots is 
\begin{multline*}
\Phi^+=\{e_i-e_j\}_{1 \leq i < j \leq m} \cup \{e_i-e_{j+m}\}_{1 \leq i \leq j \leq m} \\
\cup \{e_{i+m}-e_j\}_{1 \leq i < j \leq m} \cup \{e_{i+m}-e_{j+m}\}_{1 \leq i < j \leq m}
\end{multline*}
of cardinality $m(2m-1)$.\\
We find $m$ positive compact roots 
$$\Phi^+_c=\{e_i-e_{i+m}\}_{1 \leq i \leq m},$$
and $2m^2-2m$ positive noncompact roots
\begin{multline*}
\Phi^+_{nc}=\{e_i-e_j\}_{1 \leq i < j \leq m} \cup \{e_i-e_{j+m}\}_{1 \leq i < j \leq m} \\
\cup \{e_{i+m}-e_j\}_{1 \leq i < j \leq m} \cup \{e_{i+m}-e_{j+m}\}_{1 \leq i < j \leq m}.
\end{multline*}
The restrictions of the roots 
$$e_i-e_j,\; e_{i+m}-e_j,\; e_i-e_{j+m},\; e_{i+m}-e_{j+m}$$ 
coincide on $\mathfrak{a}$, so there are $(m^2-m)/2$ positive restricted roots $\bar{\Phi}^+=\{(e_i-e_j)_{|\mathfrak{a}}\}_{1\leq i <j \leq m}$ with  multiplicity $m_{\alpha}=4$, and $m-1$ restricted simple roots $\bar{\Delta}=\{(e_i-e_{i+1})_{|\mathfrak{a}}\}_{1\leq i \leq m-1}$. 
Consider the irreducible complex representation $\rho$ of $\mathfrak{sl}_{2m}(\C)$ over $\C^{2m}$ induced by the usual matrix-vector multiplication, and consider its restriction $\bar{\rho}=\rho_{|\bar{\mathfrak{g}}}$. Then by taking the realification of $\C^{2m}$, we can see $\bar{\rho}$ as an irreducible real representation (which coincides with the matrix/vector multiplication over $\mathbb{H}^m$; this corresponds to case (ii) in the previous discussion about irreducible real representations). The restricted weight spaces $V_{\lambda}$ for $\lambda=(e_i)_{|\mathfrak{a}}=(e_{i+m})_{|\mathfrak{a}}$ are the sum of the weight spaces for $e_i$ and $e_{i+m}$ and are generated by the vectors in $\C^{2m}$ such that the $i$-th and $(i+m)$-th component may be nonzero. The highest weight is $\lambda_1=(e_1)_{|\mathfrak{a}}=(e_{m+1})_{|\mathfrak{a}}$. 
\end{exam}

\subsubsection*{Realifications of complex Lie algebras}
The realification $\mathfrak{g}_{\R}$ of a complex Lie algebra $\mathfrak{g}$ is a real form of $\mathfrak{g} \oplus \mathfrak{g}^*$ corresponding to the involution $\sigma: (x,y) \mapsto (y,x)$.  
It can be shown \cite{VK} that the (positive) restricted root system of $\mathfrak{g}_{\R}$ coincides with the (positive) root system of $\mathfrak{g}$, but with the difference that the multiplicities $m_{\alpha}=\dim((\mathfrak{g}_{\R})_{\alpha})$ of all restricted roots are equal to $2$. 

\begin{exam}[\emph{$\mathfrak{sl}_n(\C)$ as a real Lie algebra}] \label{SLnC_real} From Example \ref{SLnC_complex}, we have
$\bar{\Delta}=\{e_i-e_{i+1}\;|\; 1 \leq i \leq n-1\}$ and $\bar{\Phi}^{+}=\{e_i-e_j\;|\; i<j\}$. The maximum weight of the realification of the complex representation considered in Example \ref{SLnC_complex} is still $e_1$. \\
\end{exam}

\subsection{Growth rate of the unit group for 
discrete subgroups of $\SL_n(\C)$} \label{volume_computation}
In this subsection we will compute the constant $T$ in Theorem \ref{Gorodnik2} in the case where $\Lambda^1$ is 
a cocompact discrete subgroup of $G=\SL_n(\C)$, $\SL_n(\R)$ or $\SL_n(\Hh)$. From Theorem \ref{Gorodnik}, we know that the growth rate of $\abs{\psi(\Lambda^1) \cap B(M)}$ depends only on the volume of the corresponding ball $B(M)$ and is the same for every discrete cocompact subgroup of $G$. Referring to the terminology of Lie algebras given in the previous subsections, we can now state a more precise version of Theorem \ref{Gorodnik2} (Theorem 7.4 in \cite{GW}). 
Let $\bar{G}$ be a connected semisimple real Lie group, $\bar{\mathfrak{g}}$ the corresponding real Lie algebra, $\bar{V}=\R^d$ a real vector space, and $\bar{\rho}: \bar{\mathfrak{g}} \to \mathfrak{gl}(\bar{V})$ an irreducible representation so that $\bar{\rho}(\bar{\mathfrak{g}})$ is identified with a subset of $M_d(\R)$. Let $\bar{\mathfrak{h}}$ be a Cartan subalgebra of $\bar{\mathfrak{g}}$ (corresponding to some choice of $\mathfrak{a}$), $\bar{\Phi}^{+}$ the restricted positive root system and $\bar{\Delta}=\{\gamma_1,\ldots,\gamma_r\}$ the corresponding set of restricted simple roots. Denote by $\lambda_1$ the highest weight of this representation.\\
 Let $\{\tilde{\beta}_1,\ldots,\tilde{\beta}_r\}$ be a basis of $\bar{\mathfrak{h}}$ such that $\gamma_i(\tilde{\beta}_j)=\delta_{i,j}$ $\forall 1 \leq i,j \leq r$. Consider the linear form $\psi=\frac{1}{2}\sum_{\gamma \in \bar{\Phi}^{+}} m_{\gamma}\gamma$, and the normalized basis $\{\beta_i\}=\left\{\frac{\tilde{\beta_i}}{2\psi(\tilde{\beta_i})}\right\}$.
 
\begin{thm}[Growth rate of units in a ball]
If the minimum $m_1=\min_{j=1,\ldots,r} \lambda_1(\beta_j)$ is achieved for only one vector $\beta_j$, then for every linear norm $\norm{\cdot}$ on $M_d(\R)$, we have 
$$\Vol(B(M)) \sim CM^{\frac{1}{m_1}},$$
for some constant $C>0$.
\end{thm}

\begin{exam}[$\SL_n(\C)$] \label{volume_SLnC}
In the case $G=\SL_n(\C)$, $\bar{\mathfrak{g}}=\mathfrak{sl}_n(\C)$. For this volume estimation, we will need to see the Lie algebra $\mathfrak{sl}_n(\C)$ as a real Lie algebra, as explained in Example \ref{SLnC_real}. As we have seen, all the restricted positive roots have multiplicity $m_{\gamma}=2$. It is not hard to see that $\tilde{\beta}_j=E_1+\ldots+E_j-jE_n$. Using the fact that $e_1 + e_2 + \ldots + e_n=0$, we get
\begin{align*}
&\psi=\sum_{i=1}^n(n-2k+1)e_k=2\sum_{k=1}^{n-1}(n-k)e_k,\\ 
&\quad \beta_j=\frac{E_1+\ldots+E_j-jE_n}{2j(2n-(j+1))} \quad \forall j\leq n-1.
\end{align*}
The minimum 
\begin{align*}
&m_1=\min_{j=1,\ldots,n-1} \lambda_1(\beta_j)=\\
&=\min_{j=1,\ldots,n-1} e_1\left(\frac{E_1+\ldots+E_j-jE_n}{2j(2n-(j+1))}\right)
\end{align*}
is achieved for the unique value $j=n-1$ and is equal to $\frac{1}{2n(n-1)}$. So $T=2n(n-1)$.  
\end{exam}

\begin{exam}[$\SL_n(\R)$] \label{volume_SLnR}
The case $G=\SL_n(\R)$, $\bar{\mathfrak{g}}=\mathfrak{sl}_n(\R)$ is similar to the previous one except for the fact that $m_{\gamma}=1$ (see Example \ref{SLnR}). Consequently, we obtain $T=n(n-1)$.
\end{exam}

\begin{exam}[$\SL_m(\Hh)$] \label{volume_SLnH}
For $G=\SL_m(\Hh)$, $\bar{\mathfrak{g}}=\mathfrak{sl}_m(\Hh)$, we refer to Example \ref{SLnH}.
We can choose the dual basis
$$\{\tilde{\beta_j}\}=\{E_1+E_{m+1}+\ldots+E_j+E_{m+j}-jE_m-jE_{2m}\}_{1 \leq j \leq m-1}$$ 
and the linear form
$$\psi=2\sum_{k=1}^{m-1}(m-k)(e_k+e_{k+m}),$$
and similarly to before, we find $T=4m(m-1)$. 
\end{exam}
\end{appendices}

\section*{Acknowledgement}

The authors would like to thank Alexander Gorodnik  for answering questions concerning point counting in Lie groups and the reviewers for their hard work that has benefited us greatly.

\begin{IEEEbiographynophoto}{Roope Vehkalahti}
received the M.Sc. and Ph.D. degrees from the University of Turku, Finland,
in 2003 and 2008, respectively, both in pure mathematics.

Since September 2003, he has been with the Department of
Mathematics, University of Turku, Finland.  In 2011-2012 he was visiting Swiss Federal Institute of Technology, Lausanne (EPFL).  His  research interest include  applications of algebra and number theory to information theory.
\end{IEEEbiographynophoto}

\begin{IEEEbiographynophoto}{Hsiao-feng (Francis) Lu}
(S'98-M'04-SM'12) received the B.S. 
degree from Tatung University, Taipei,
Taiwan, in 1994, and the M.S. and Ph.D. degrees from 
University of Southern California (USC), Los Angeles, in 1999 and
2003, respectively, all in electrical engineering.

He was a postdoctoral research fellow at University of Waterloo, ON,
Canada, during 2003-2004. In February 2004, he joined the
Department of Communications Engineering, National Chung-Cheng
University, Chiayi, Taiwan, where he was promoted to Associate 
Professor in August 2007. Since August 2008, he has been with the Department of
Electrical Engineering, National Chiao Tung University, Hsinchu,
Taiwan, where he is currently a Full Professor. His research is in the area of space-time codes, MIMO systems,
error correcting codes, wireless communication, optical fiber
communication, and multi-user detection. He is an Associate Editor of 
{\em IEEE Transactions on Vehicular Technology}.

Dr. Lu is a recipient of several research awards, including the 2006 
IEEE
Information Society Taipei Chapter and IEEE Communications Society
Taipei/Tainan Chapter Best Paper Award for Young Scholars, the 2007
Wu Da You Memorial award from Taiwan National Science Council, the
2007 IEEE Communication Society  Asia Pacific Outstanding Young
Researchers Award, and the 2008 Academia Sinica Research Award for
Junior Research Investigators.
\end{IEEEbiographynophoto}

\begin{IEEEbiographynophoto}{Laura Luzzi} received the degree (Laurea) in
Mathematics from the University of Pisa, Italy, in 2003 and the Ph.D.
degree in Mathematics for Technology and Industrial Applications from
Scuola Normale Superiore, Pisa, Italy, in 2007. From 2007 to 2012 she held
postdoctoral positions in T\'el\'ecom-ParisTech and Sup\'elec, France, and
a Marie Curie IEF Fellowship at Imperial College London, United Kingdom.
She is currently an Assistant Professor at ENSEA de Cergy, Cergy-Pontoise,
France, and a researcher at Laboratoire ETIS (ENSEA - Universit\'e de
Cergy-Pontoise- CNRS).\\
Her research interests include algebraic space-time coding and decoding
for wireless communications and physical layer security.
\end{IEEEbiographynophoto}

\end{document}